\documentclass[preprint]{aastex}
\usepackage{natbib}
\begin{document}

\title{Mapping H-band Scattered Light Emission in the Mysterious SR21 Transitional Disk \thanks{Based on data collected at Subaru Telescope, which is operated by the National Astronomical Observatory of Japan.}}

\author{Katherine B. Follette\altaffilmark{1}, Motohide Tamura\altaffilmark{2,3}, Jun Hashimoto\altaffilmark{2}, Barbara Whitney\altaffilmark{4}, Carol Grady\altaffilmark{5}, Laird Close\altaffilmark{1}, Sean M. Andrews\altaffilmark{6}, Jungmi Kwon\altaffilmark{3,2}, John Wisniewski\altaffilmark{7}, Timothy D. Brandt\altaffilmark{8}, Satoshi Mayama\altaffilmark{9}, Ryo Kandori\altaffilmark{2}, Ruobing Dong\altaffilmark{8}, Lyu Abe\altaffilmark{10}, Wolfgang Brandner\altaffilmark{11}, Joseph Carson\altaffilmark{12}, Thayne Currie\altaffilmark{13}, Sebastian E. Egner\altaffilmark{14}, Markus Feldt\altaffilmark{11}, Miwa Goto\altaffilmark{15}, Olivier Guyon\altaffilmark{14}, Yutaka Hayano\altaffilmark{14}, Masahiko Hayashi\altaffilmark{16}, Saeko Hayashi\altaffilmark{14}, Thomas Henning\altaffilmark{11}, Klaus Hodapp\altaffilmark{17}, Miki Ishii\altaffilmark{14}, Masanori Iye\altaffilmark{2}, Markus Janson\altaffilmark{8}, Gillian R. Knapp\altaffilmark{8}, Tomoyuki Kudo\altaffilmark{14}, Nobuhiko Kusakabe\altaffilmark{2}, Masayuki Kuzuhara\altaffilmark{2}, Michael W. McElwain\altaffilmark{18}, Taro Matsuo\altaffilmark{19}, Shoken Miyama\altaffilmark{20}, Jun-Ichi Morino\altaffilmark{2}, Amaya Moro-Martin\altaffilmark{21}, Tetsuo Nishimura\altaffilmark{14}, Tae-Soo Pyo\altaffilmark{14}, Eugene Serabyn\altaffilmark{22}, Hiroshi Suto\altaffilmark{2}, Ryuji Suzuki\altaffilmark{23}, Michihiro Takami\altaffilmark{24}, Naruhisa Takato\altaffilmark{14}, Hiroshi Terada\altaffilmark{14}, Christian Thalmann\altaffilmark{25}, Daigo Tomono\altaffilmark{14}, Edwin L. Turner\altaffilmark{8,26} Makoto Watanabe\altaffilmark{27}, Toru Yamada\altaffilmark{28}, Hideki Takami\altaffilmark{14}, Tomonori Usuda\altaffilmark{14}}
\altaffiltext {1}{Steward Observatory, The University of Arizona, 933 N Cherry Ave, Tucson, AZ 85721, USA}
\altaffiltext{2}{National Astronomical Observatory of Japan, 2-21-1 Osawa, Mitaka, Tokyo 181-8588, Japan}
\altaffiltext{3}{Department of Astronomical Science, Graduate University for Advanced Studies (Sokendai), Tokyo 181-8588, Japan}
\altaffiltext{4}{Astronomy Department, University of Wisconsin-Madison, 475 N. Charter Street, Madison, WI 53706, USA}
\email{kfollette@as.arizona.edu}
\altaffiltext {5}{Eureka Scientific, 2452 Delmer, Suite 100, Oakland CA 96002, USA}
\altaffiltext{6}{Harvard-Smithsonian Center for Astrophysics, 60 Garden Street, Cambridge, MA, 02138, USA}
\altaffiltext{7}{H.L. Dodge Department of Physics and Astronomy, University of Oklahoma, 440 W Brooks St Norman, OK 73019, USA}
\altaffiltext{8}{Department of Astrophysical Sciences, Princeton University, NJ 08544, USA}
\altaffiltext{9}{The Center for the Promotion of Integrated Sciences, The Graduate University for Advanced Studies (Sokendai), Shonan International Village, Hayama-cho, Miura-gun, Kanagawa 240-0193, Japan}
\altaffiltext{10}{Laboratoire Lagrange, UMR7293, Universit\'{e} de Nice-Sophia Antipolis, CNRS, Observatoire de la C\^{o}te d'Azur, 06300 Nice, France}
\altaffiltext{11}{Max Planck Institute for Astronomy, K\"{o}nigstuhl 17, 69117 Heidelberg, Germany}
\altaffiltext{12}{Department of Physics and Astronomy, College of Charleston, 58 Coming St., Charleston, SC 29424, USA}
\altaffiltext{13}{Department of Astronomy and Astrophysics, University of Toronto, 50 St. George Street M5S 3H4, Toronto Ontario Canada}
\altaffiltext{14}{Subaru Telescope, 650 North AÕohoku Place, Hilo, HI 96720, USA}
\altaffiltext{15}{Universitats-Sternwarte Munchen, Ludwig-Maximilians-Universitat, Scheinerstr. 1, 81679 Munchen, Germany}
\altaffiltext{16}{Department of Astronomy, The University of Tokyo, Hongo 7-3-1, Bunkyo-ku, Tokyo 113-0033, Japan}
\altaffiltext{17}{Institute for Astronomy, University of Hawaii, 640 North A'ohoku Place, Hilo, HI 96720, USA}
\altaffiltext{18}{ExoPlanets and Stellar Astrophysics Laboratory, Code 667, Goddard Space Flight Center, Greenbelt, MD 20771 USA}
\altaffiltext{19}{Department of Astronomy, Kyoto University, Kitashirakawa-Oiwake-cho, Sakyo-ku, Kyoto, 606-8502, Japan}
\altaffiltext{20}{Office of the President, Hiroshima University, 1-3-2 Kagamiyama, Higashi-Hiroshima, 739-8511, Japan}
\altaffiltext{21}{Departamento de Astrof\'{i}sica, CAB (INTA-CSIC), Instituto Nacional de T\'{e}cnica Aeroespacial, Torrej\'{o}nde Ardoz, 28850, Madrid, Spain}
\altaffiltext{22}{Jet Propulsion Laboratory, California Institute of Technology, Pasadena, CA, 91109 USA}
\altaffiltext{23}{TMT Observatory Corporation, 1111 South Arroyo Parkway, Pasadena, CA 91105, USA}
\altaffiltext{24}{Institute of Astronomy and Astrophysics, Academia Sinica, P.O. Box 23-141, Taipei 106, Taiwan}
\altaffiltext{25}{Astronomical Institute ``Anton Pannekoek'', University of Amsterdam, Science Park 904, 1098 XH Amsterdam, The Netherlands}
\altaffiltext{26}{Kavli Institute for the Physics and Mathematics of the Universe, The University of Tokyo, Kashiwa 277-8568, Japan}
\altaffiltext{27}{Department of Cosmosciences, Hokkaido University, Sapporo 060-0810, Japan}
\altaffiltext{28}{Astronomical Institute, Tohoku University, Aoba, Sendai 980-8578, Japan}

%\maketitle

\begin{abstract}
We present the first near infrared (NIR) spatially resolved images of the circumstellar transitional disk around SR21. These images were obtained with the Subaru HiCIAO camera, adaptive optics and the polarized differential imaging (PDI) technique. We resolve the disk in scattered light at H-band for stellocentric 0$\farcs$1$\leqslant$$r$$\leqslant$0$\farcs$6 (12$\lesssim$$r$$\lesssim$75AU). We compare our results with previously published spatially-resolved 880$\mu$m continuum Submillimeter Array (SMA) images that show an inner $r$$\lesssim$36AU cavity in SR21. Radiative transfer models reveal that the large disk depletion factor invoked to explain SR21's sub-mm cavity cannot be ``universal'' for all grain sizes. Even significantly more moderate depletions ($\delta$=0.1, 0.01 relative to an undepleted disk) than those that reproduce the sub-mm cavity ($\delta$$\sim$10$^{-6}$) are inconsistent with our H-band images when they are assumed to carry over to small grains, suggesting that surface grains scattering in the NIR either survive or are generated by whatever mechanism is clearing the disk midplane. In fact, the radial polarized intensity profile of our H-band observations is smooth and steeply inwardly-increasing ($r$$^{-3}$), with no evidence of a break at the 36AU sub-mm cavity wall. We hypothesize that this profile is dominated by an optically thin disk envelope or atmosphere component. We also discuss the compatibility of our data with the previously postulated existence of a sub-stellar companion to SR21 at $r$$\sim$10-20AU, and find that we can neither exclude nor verify this scenario. This study demonstrates the power of multiwavelength imaging of transitional disks to inform modeling efforts, including the debate over precisely what physical mechanism is responsible for clearing these disks of their large midplane grains.   
\end{abstract}

\section{Introduction}
Transitional disks are of particular interest to the astronomical community because they have features intermediate between optically thick gas-rich protoplanetary disks and optically thin dusty debris disks, suggesting that they may be an intermediate evolutionary step.  Transitional disks were first classified based on their Spectral Energy Distributions (SEDs) as objects with significant mid infrared excess emission indicative of starlight reprocessed by a circumstellar disk, but a lack of strong near infrared (NIR, $\sim$1-5$\mu$m) excess. This NIR deficit implies a dearth of the very warmest grains in the inner disk \citep{Strom:1989,Skrutskie:1990}. Spatially-resolved sub-mm imaging of a number of transitional disks has revealed that many of them are indeed cleared of large thermally-emitting dust grains out to 10AU or more, well beyond the dust sublimation radius, thereby suggesting that an additional clearing mechanism may be at work \citep[referred to as A11 throughout the remainder of this paper]{Andrews:2009, Hughes:2009a, Brown:2009,Andrews:2011}.

The precise mechanism by which these clearings are made is widely debated in the literature, with several viable alternatives put forward, including clearing by massive forming protoplanets \citep[e.g.][]{Dodson-Robinson:2011, Zhu:2011}, clearing by stellar mass companions \citep[e.g.,][]{Ireland:2008, Jensen:1997}, photoevaporation \citep[e.g.][]{Owen:2011, Clarke:2001, Hollenbach:1994} and grain growth \citep[e.g.,][]{Birnstiel:2012,Dullemond:2005}.  Each theoretical explanation for the observed presence of sub-mm disk cavities makes itÕs own predictions for clearing timescales, efficiencies and radii. However, the small amount of available spatially-resolved multiwavelength data on transitional disks has limited the ability of modelers to test these theories against observation. As such data accumulate, they reveal an increasingly complex picture of the ``cleared" regions. In particular, cavities seen at long wavelengths do not necessarily manifest themselves in NIR scattered light \citep[e.g.this work, ][]{Dong:2012}. Spectroscopic studies also reveal a variety of gaseous species in the inner regions of transitional disks \citep{Salyk:2009,Pontoppidan:2008}. As large dust grains, small dust grains and gas all appear to respond quite differently to the mechanism or mechanisms that create transitional disks, a thorough understanding of their clearings can only be grasped through accumulation of a statistically significant sample of multiwavelength data on transitional disks.  

Thus far, the most success in spatially resolving transitional disks has been achieved in the sub-mm regime (e.g., A11), with a small number of additional studies using mid-infrared (MIR, 10-40$\mu$m) aperture masking \citep[e.g.][hereafter E09]{Eisner:2009}, NIR spectroastrometry \citep[e.g.][hereafter P08]{Pontoppidan:2008} and MIR interferometry \citep[e.g.][]{Ratzka:2007} to infer inner disk structure. Efforts are currently underway to image a sample of transitional disks at shorter wavelengths, including the Strategic Exploration of Exoplanets and Disks with Subaru \citep[SEEDS;][]{Tamura:2009} campaign. 

Despite initial success at resolving the rim of the LkCa 15 cavity in scattered light \citep{Thalmann:2010} with a Locally Optimized Combination of Images (LOCI) method, additional transitional disks imaged by SEEDS are revealing that these cavities, including that of LkCa 15, are more elusive when viewed in polarized light \citep[Wisniewski et al. 2013, in prep,]{Hashimoto:2012,Muto:2012,Dong:2012,Mayama:2012,Grady:2013}. On the surface, this suggests that NIR scattering material is present, and even abundant, inside the cavities. Although sub-mm studies are generally revealing that transitional objects are indeed cleared or depleted of large grain material inside of the cavity radii inferred from their SEDs, NIR `clearings` in transitional disks may not be as ubiquitous nor as fully cleared as their sub-mm cousins.  In fact, it appears that some transitional disks contain abundant small grain material inside of observed sub-mm cavities. 

Polarized NIR scattered light and sub-mm imaging, though both sensitive to circumstellar dust, probe very different regimes in the disk, both in terms of density and height above the midplane. The sub-mm emission from transitional disks is thermal, tracing mainly mm-sized grains that are concentrated (``settled") toward the disk midplane \citep[e.g.,][]{Dullemond:2004,Brauer:2008,DAlessio:2006}. These large ($a$$>$10$\mu$m) grains are very important for understanding the large scale structure of the disk, as they house most of the disk's dust mass and their coagulation drives the process of planet formation under the core accretion scenario  \citep[e.g.,][]{Pollack:1996}. The density and temperature of the grains at the disk midplane also determine the size and shape of the remainder of the disk. 

At NIR wavelengths, however, it is unclear whether scattered light emission originates from an optically thick small grain disk surface, or from more diffuse material in or above the disk. This uncertainty that is compounded by the small number of spatially resolved disks in the literature. It is unknown whether disks with large grain clearings in the sub-mm are also cleared of small grains, nor is it understood what proportion of the disk mass lies in the small grain population. The fact that we see axisymmetric polarization structure in the NIR from transitional objects reveals that they house circumstellar disks with some small grain material \citep{Whitney:1992}. However, very little dust mass is needed to scatter at H-band, so it is unclear how much mass resides in this portion of the disk and whether it is optically thin or optically thick in this wavelength regime. If transitional disks are optically thick in the NIR, as is the case for protoplanetary disks, then this scattered emission will probe only the $\tau$=1 surface. Small grains are believed to be vertically extended by a factor of two or more relative to larger grains \citep{Dullemond:2004, Brauer:2008}, so NIR scattering in an optically thick disk should occur well above the disk midplane. It is unclear whether or not clearing mechanisms acting on the large grains at the disk midplane would have an observable effect on the NIR emission under this scenario.

Conversely, if transitional disks are optically thin in this wavelength regime, as is the case for debris disks, then scattered light emission probes the entire vertical extent of the disk. In this case, scattered light emission originates from small grains throughout the vertical extent of the disk, down to and including the disk midplane. Observations of the same disk at both sub-mm and NIR wavelengths  are compelling in that they allow for comparison of the settled and upper surface layers of the disk, the large (mm-sized) and small ($\mu$m-sized) grains, separately. A good disk model should thus be able to self-consistently reproduce observations at both wavelengths simultaneously, as well as any additional multiwavelength data. Our efforts to do just that for the SR21 disk are described in detail in Section 4.  Section 2 describes our data collection methodology, and our result are summarized in Section 3. 

\subsection{Observational History of SR21}
SR21A is a 2.5M$_{\sun}$, $\sim$1 Myr old Class II Young Stellar Object (YSO) in the Ophiucus Star Forming Region. Recent astrometric observations suggest that the distance to the Ophiuchus star forming region is 119$\pm$6 pc \citep{Lombardi:2008}. However, we adopt a distance of 135pc \citep{Mamajek:2008} in order to remain consistent with previous studies. SR21 has been classified as a 6$\farcs$7 binary \citep{Barsony:2003}; however, \citet{Prato:2003} concluded that the two components were not coeval, and that the B component does not exhibit infrared excess emission and is therefore unlikely to host a circumstellar disk of its own. The A and B components also have discrepant proper motions \citep{Roeser:2010}, so the``binary" does not appear to be a bound, coeval system. We will therefore refer to SR21A as simply SR21 throughout the remainder of this paper. 

SR21 was first identified as a transitional disk by \citet{Brown:2007}, who inferred a 0.45-18AU gap in the disk based on a fit to the $\sim$1-100$\mu$m SED. In order to explain the NIR excess seen at $<$5$\mu$m, they invoked a small uncleared region in the inner disk from 0.25-0.45AU. Since this initial identification, a variety of puzzling and seemingly contradictory data have emerged that continue to favor a disk that is not simply cleared of all material in this 0.45-18AU region.  

The first evidence of a lack of complete clearing in the inner disk came from CO rovibrational spectroastrometry done with CRIRES on the VLT by P08, who observed molecular gas emission originating from a region well inside of the dust gap inferred by Brown et al. They concluded that this emission arose from a narrow 7-7.5AU ring of molecular gas. It is not surprising to have found gas emission inside of the dust gap, as SR21 may still be accreting, albeit weakly \citep[$<$10$^{-8.84}$M$_{\sun}$/yr, a relatively high upper limit,][]{Natta:2006}. However, the physical mechanism causing an apparently truncated ring of molecular gas at this radius is still an open question. P08 suggest that the inner 7AU cutoff corresponds to a physical truncation of the disk, while a rapid drop in gas temperature outward of 7AU, perhaps due to a thick disk component in the region, causes the CO flux to quickly drop below the detection threshold. This explains the observed truncation of CO emission at 7.5AU without requiring that the disk material be physically truncated at this radius. 

 E09 also postulated that there is emission from within the $\sim$18AU `cavity' based on MIR (8.8 and 11.6$\mu$m) aperture masking data taken with TReCS on Gemini South. They put forward an alternate interpretation to \citet{Brown:2007}, capable of fitting both the SR21 SED and their visibility data. They suggest that the disk that is completely cleared within 10AU, but with a warm companion at 10-18AU whose circum(sub)stellar material creates the NIR excess seen in the SED.  

The first spatially resolved imaging study of SR21 was done at 340GHz (880$\mu$m) and was published by \citet{Brown:2009} among a sample of four disks with large central clearings. Followup data, also at 340GHz, were presented in \citet{Andrews:2009}.  We base our sub-mm modeling efforts on A11, a detailed imaging and modeling study of 12 transitional disks, including SR21. Using two-dimensional Monte Carlo radiative transfer modeling to fit the sub-mm data and the infrared SED, A11 concluded that the 36AU sub-mm central cavity in SR21 was depleted in material by a factor of $\sim$10$^{-6}$ relative to the outer disk. 

As multiwavelength data accumulate on SR21, it is clear that its structure, particularly in the region interior to the 36AU sub-mm clearing, cannot be understood through SED modeling alone because there are several degenerate solutions capable of explaining it \citep{Brown:2007,Eisner:2009}. Spatially resolved images are needed at multiple wavelengths, probing multiple regions of the disk, before good theoretical constraints can be put on the mechanism clearing the interior of SR21. In the next section, we describe new spatially resolved NIR scattered light imaging, which contributes further to the puzzle of the SR21 ``cavity". 

\section{Observations and Data Reduction}
Polarized Differential Imaging (PDI) of SR21 was done in H-band (1.6$\mu$m) on 2011 May 22 with the high-contrast imager HiCIAO \citep{Tamura:2006, Hodapp:2006} and adaptive optics system AO188 \citep{Minowa:2010} on the 8.2m Subaru telescope on Mauna Kea. These observations were conducted as part of the larger SEEDS survey, which was begun in October 2009 and is currently entering its third year. 
	
The data were taken in ``qPDI'' mode in which a double Wollaston prism is used to split the beam into four 512x512 pixel channels (0$\farcs$0095/pixel), two each of o and e-polarizations. The splitting of each polarization state into two separate channels reduces saturation effects on the Hawaii 2RG detector. In order to obtain full polarization coverage and minimize artifacts, a half waveplate was rotated to four different angular positions (0$^{\circ}$, 45$^{\circ}$, 22$\fdg$5 and 67$\fdg$5). This cycle was repeated 18 times on SR21, with a 15 second exposure per waveplate position, for a total of 18 minutes of integration time. 
	
Each image was bias subtracted, flat fielded and bad pixels removed in the standard manner for infrared data analysis using custom IDL scripts. Pinhole data were then used to create a distortion solution for each channel, which was applied before the images were rotated to a common on-sky geometry. All of the 15 second science exposures of SR21 were saturated to a radius of 6-8 pixels. For this reason, all subsequent combinations have had an inner $r$=8 pixel mask applied. 
	 
Each individual channel was then aligned using Fourier cross-correlation. These aligned images were combined to create Stokes Q and U images using standard differential polarimetry methods, namely adding together each set of two identically polarized channels and then subtracting these combinations from one another. The four channels were also added directly together to form a total intensity image.

To estimate the performance of the Subaru AO system during our observations, we used two unsaturated images of the Point Spread Function (PSF) reference star HD 148212, and 12 and 10 unsaturated images of SR21 taken before and after the main SR21 data set respectively. The average Strehl Ratio of the HD 148212 images is 0.27, and the average Strehl Ratios of the SR21 before and after images are 0.22 and 0.20 respectively. The standard deviations of the Strehl Ratios in the two SR21 data sets are 0.018 and 0.016 respectively, indicating that AO correction was maintained steadily.

\subsection{Polarized Intensity Images and Polarization Vectors}

The polarized intensity image shown in Figure 1 is a median combination of individual Polarized Intensity (PI) images for each of the 18 waveplate cycles, computed via $PI=\sqrt{Q^{2} + U^{2}}$. Polarization angles were calculated for each image according to the formula  $\theta_{p}= 0.5 tan^{-1} (U/Q)$ and median combined. These are shown overplotted on the PI image in five pixel bins in the lefthand panel of Figure 2. The length of each vector has been scaled according to the magnitude of the polarized intensity in that bin. A first order instrumental polarization correction has been applied, as described in detail in \citet{Hashimoto:2011}. 

As has been demonstrated by other groups \citep{Close:1997, Oppenheimer:2008, Apai:2004, Quanz:2011}, PI images are a powerful data product for studying circumstellar disks. Unlike most other varieties of disk observations, PI data are generally spared the difficult and imperfect task of PSF subtraction under the assumption that direct starlight is randomly polarized, while light scattered off of a circumstellar disk shows a preferred polarization direction according to the scattering location in the disk. Specifically, the scattering process will preferentially scatter light polarized perpendicular to the direction to the illuminating source, while light coming directly from the star will be unpolarized and randomly oriented. Extended polarized emission is therefore unlikely to occur in systems without scattered light disks, and light scattered from a disk should show a centrosymmetric pattern of polarization vectors. 

Although direct starlight is randomly polarized when it reaches the telescope, multiple reflective surfaces in the HiCIAO optical path are capable of polarizing some fraction of this incoming light. This phenomenon is referred to as instrumental polarization (IP). In order to estimate the quality of IP removal via standard techniques in our data, we reduced our data on the PSF star HD 148212 using the same methodology as SR21, including a first-order instrumental polarization correction. HD 148212 is a G2V star at d=105$\pm$15pc and $\sim$10' from SR21. It was observed just before SR21 on the same night and in the same observing mode, with 17 cycles of waveplate rotations and 5 seconds per exposure, for a total of 5.7 minutes of exposure time. As shown in the righthand panel of Figure 2, no polarized emission was observed beyond $r$$\gtrsim$0$\farcs$2 in the initial first order IP corrected HD 148212 polarized intensity image. Although the polarization signal inside of this radius is small compared to the Polarized Intensity of SR21 in the same region, it does show the characteristic four-lobed astigmatic pattern of the HiCIAO PSF. Additionally, while polarization vectors for HD 148212 do not show centrosymmetry, they do show some alignment along the vertical axis of the instrument. These are both hints that there is a residual uncorrected instrumental polarization signature, and further correction is warranted. 

To first order, we can approximate the residual instrumental polarization by measuring the average polarization strength and direction of the signal in the HD 148212 polarized intensity image. We assume that HD 148212 is an ordinary diskless unpolarized star, and that all of its extended emission is due to residual instrumental polarization or photon noise. We construct an artificial ``halo" by multiplying the average strength (1.1\%) in the HD 148212 halo into the raw intensity image, then creating artificial ``halo Q" and ``halo U" images from it using the measured average polarization angle (-11$^{\circ}$). We combine these synthetic Q and U images to create a ``halo PI" image and subtract it from the original HD 148212 PI image. We also subtract the ``halo Q" and ``halo U" images from the raw Q and U images, and use the resulting images to calculate corrected polarization vectors, which we overplot on the halo-subracted PI image for HD 148212 in the righthand panel of Figure 2 . The extended polarized intensity emission from the PSF star is reduced to $r$$<$0$\farcs$1 through this process, and the resulting vectors are randomized, no longer showing a preferred polarization direction. We are therefore confident that this provides a reasonable approximation of the residual instrumental polarization.

The need to apply this same ``halo removal" procedure to the SR21 PI data can be seen in the Figure 2 SR21 polarization map. Although the SR21 vector pattern is clearly centrosymmetric along and around the disk major axis even in the uncorrected image, it deviates somewhat from centrosymmetry along the same instrumental axis as the aligned vectors in the HD 148212 image (approximately the minor axis of the disk). Since not all SR21 polarized emission is spurious, approximating the strength and direction of the halo is nontrivial in this case.  We used a region with low intrinsic disk emission to construct a ``halo" for subtraction, namely 100 pixel (0$\farcs$95) wide regions centered on the minor axis at 30 pixels$<$$r$$<$50 pixels (0$\farcs$285$<$$r$$<$0$\farcs$475) on either side of the disk.  Following the same procedure as with HD 148212, we created polarized intensity ``halo" Q, U and PI images from the polarization direction (-8.8$^{\circ}$) and strength (1.02\%) measured in this region and subtracted. 

Subtraction of this first-order approximation reveals a more centrosymmetric pattern, as evidenced in the lefthand panel of Figure 3. However, it also introduces significant PSF artifacts to the polarized intensity map, as can be seen in the background polarized intensity contours of Figure 3. For this reason, we have chosen to model the ``raw" unsubtracted radial profile in Section 4, noting, however, that some of the emission interior to 0$\farcs$15 is likely to be contaminated by a small degree of instrumentally polarized stellar flux. The polarized halo strength of 1.02\% is an order of magnitude smaller than our lower limit for the intrinsic SR21 polarization (10\%), so this contamination is minimal.  

Despite its crudeness, the ``halo subtraction" method highlights an interesting feature of the SR21 polarization vector structure, which is that several regions in the disk with high signal (notably the northeast and southwest quadrants) show deviations from centrosymmetry even after halo subtraction. This may indicate that there is unresolved structure in the disk at these locations. Interestingly, the deviation in the northeast is consistent with the location of the companion inferred by E09, which could be explained if the companion had an accretion disk of its own, with its own polarization structure superimposed on that of the SR21 disk. 

\subsection{Intensity Images}

Intensity images were also constructed from the PDI observations of SR21 by cross correlating, centroiding and aligning the images in each of the four channels at each of the four waveplate (0$^{\circ}$, 45$^{\circ}$, 22$\fdg$5 and 67$\fdg$5) positions for each of the 18 observation cycles. The four channels at each waveplate position were summed for each cycle, and the resulting 18 images were then median combined to create an SR21 intensity image. 

Unlike polarized intensity images, total intensity images are dominated by stellar flux, and PSF subtraction is necessary. For this purpose, we created intensity images of HD 148212 using the same methodology described above. The final reduced (but unsubtracted) median combinations of the SR21 and HD 148212 total intensity images are shown in Figure 4. Both reveal significant uncorrected astigmatism, as evidenced by the four-lobed pattern centered on the star. The SR21 disk is not visible in the unsubtracted total intensity image. The PSF subtraction process is described in detail in Section 3.3.

\section{Results and Analysis}

Our principle observational results are the final polarized intensity (Figure 1) and total intensity (Figure 4) images and polarization vectors (Figures 2 and 3). These images were analyzed in order to estimate the extent, inclination, major axis orientation and polarization percentage of the SR21 disk. The derivation of each of these properties is described in detail in this section. 

\subsection{Polarized Intensity Isophotes}
We began by fitting elliptical isophotes to our PI data in order to provide our own measure of the inclination and major axis of the SR21 disk, independent of the values in the literature, which were determined at other wavelengths. These elliptical fits are shown overplotted on Figure 5 in white, beginning at a radius of 10 pixels from the center (well outside the saturation radius) and extending outward to 50 pixels, where the PI flux approaches the noise level. Errors in isophotal fits were determined by summing in quadrature the intrinsic error estimation in the IRAF\footnote{IRAF is distributed by the National Optical Astronomy Observatory, which is operated by the Association of Universities for Research in Astronomy (AURA) under cooperative agreement with the National Science Foundation.} ellipse procedure, which was used to fit the isophotes, and the standard deviation of the parameter across the image. 

The average ellipticity of these isophotes suggests a disk inclination of 14$\pm$2$^{\circ}$ (where 0$^{\circ}$ would be a face-on geometry). This is smaller than the value reported by A11; however, their estimate of 22$^{\circ}$ was made with the caveat that the sub-mm ring seen in SR21 is very narrow, and the geometry is not well constrained by their observations. The H-band emission is significantly more spatially extended; therefore, we adopt 14$^{\circ}$ as the preferred value in our modeling efforts, which are described in section 4.  There is no statistically significant deviation in our calculated inclination value between the inner and outer disk of SR21.

There is a clear counterclockwise progression from the inner to the outer disk in the major axis of the elliptical isophotal fits (overplotted in white in Figure 5). The average orientation of the major axis over the entire disk is 86$\pm$11$^{\circ}$. The large uncertainty is due to the fact that the PA migrates from inner to outer disk. This happens relatively smoothly, but with a pronounced jump between 17 and 20 pixels. Inside of this radius, which is of particular interest because it corresponds to the location of E09's hypothesized companion, the average major axis position angle is 76$\pm$8$^{\circ}$ (measured East of North), while outside of this radius it is 96$\pm$7$^{\circ}$. This shift in the average major axis PA of $\sim$20$^{\circ}$ between these two regions, and of 42$^{\circ}$ in total between inner (68$^{\circ}$) and outermost (110$^{\circ}$) isophotes, suggests that the SR21 disk may contain unresolved structures such as a warp or a spiral arm. Some of this PA migration might be explained by instrumentally polarized emission with the shape of the HiCIAO PSF, particularly because the most pronounced turnover occurs at approximately the same radius as the extent of the HD 148212 halo. However, the observed SR21 isophote major axis migration continues well beyond that radius, and we therefore conclude that it is a real feature of the data. 

We combine our results with previous estimates of the major axis PA from the literature in Table 1. The A11 estimate for the major axis PA of 100$^{\circ}$ (overplotted on Figure 5 in red) is consistent with our measurement of the PA of the outer disk.   We find that previous estimates of the PA of the disk are strongly dependent on the radius being probed, and our observation of a counterclockwise trend with radius fits well into the broader picture.  This suggests that there is unresolved non-axisymmetric structure in the SR21 disk, and provides further confirmation that the trend we observe is not caused purely by residual PSF-shaped instrumental polarization signal. 

\subsection{Radial Brightness Profiles and Asymmetries}

We used the final median combination of unsaturated total intensity images and the published 2MASS photometry values of SR21 and HD 148212 to establish a flux conversion for our data. In doing so, we make the assumption that the H-band magnitudes of both objects remained steady in the decade between 2MASS and HiCIAO observations. H-band emission of T Tauri stars has been shown to be variable \citep{Carpenter:2002}, however the The ADU-to-Jansky conversion factor calculated using the HD 148212 images differs by only 1.6\% from that calculated using the SR21 images; therefore, we believe that this is a robust assumption and have adopted the average of the two values as our flux conversion factor. We have applied a software mask to the saturated region interior to $r$=8 pixels; however, astigmatic PSF artifacts extend out to $r$$\sim$13 pixels (0$\farcs$13, $\sim$18AU). Our data are therefore most robust at 18AU$\lesssim$$r$$\lesssim$80AU. 

Computed flux-converted radial polarized intensity profiles for SR21 are shown in Figure 6. Brightness profiles are shown for both the azimuthally-averaged case (filled black points) and for the case where we limit analysis to a region 11 pixels wide centered on the disk major axis (open red points). We also over plot (open blue points) the radial profile of a PSF-subtracted Polarized Intensity image. Although PSF subtraction is nominally unnecessary for polarized intensity images, our analysis in section 2.1 revealed that residual instrumental polarization contributes at a low level. We therefore utilized the polarized intensity image of the PSF star HD 148212, and scaled it such that the noise in the background region was equivalent to that of SR21 and subtracted. This provided a subtraction with fewer evident PSF artifacts than our earlier ``halo subtracted" PI image and serves to verify that the slope and steepness of the radial polarized intensity profile is insensitive to contamination by residual instrumental polarization. 

In all three cases, we plotted the profile in five pixel radial bins excluding the innermost two bins (so interior to 0$\farcs$1), which have pixels inside the saturation radius. This binning was chosen to match the diffraction limit of 0$\farcs$05 at H-band. The brightness profile along the major axis is systematically higher than the azimuthally-averaged profile, as is to be expected based on the increased brightness of the disk in this region; however, the shapes are virtually identical. The PSF subtracted profile is systematically lower than that of the unsubtracted profile, as is to be expected. Unlike the other two profiles, the PSF-subtracted profile does not flatten in the outer disk, suggesting that this trend may be caused by residual instrumental polarization in the unsubtracted profiles.  We have elected to fit the azimuthally-averaged profile of the unsubtracted image in our modeling, which is described in Section 4. 

These radial PI profiles reveal that H-band scattered light emission exists in the SR21 disk out to a distance of at least $\sim$0$\farcs$6, which is comparable to the disk extent in the A11 SMA images. This corresponds to a physical scale of $\sim$80AU, or more than twice the radius of the observed sub-mm cavity.  The excess in the NE quadrant of the disk, which could be due to the contribution of a companion as discussed in Section 3.3, is outside of the region used to create the major axis radial profile, and is smoothed out in the azimuthally averaged profile, therefore does not have any noticeable effect on either of our derived radial profiles.

Errors in the radial polarized intensity measurements were estimated using both the background photon noise in the four individual qPDI channels of the SR21 raw images and using the observations of the PSF star HD 148212. The residual polarized intensity flux surrounding HD 148212 was summed in the same manner as the SR21 radial profile and is comparable to the photon noise. The error bars in Figure 6 and all of the radial profile plots shown in the modeling section correspond to the photon noise.  

A canonical $r$$^{-2}$ power law for the surface brightness of a flared disk is shown overplotted in green on all radial profiles and is a poor fit to the data. The best fit radial profile to the data is of the form $r$$^{-3}$. Although there is some indication that it may flatten beyond 0$\farcs$45, this trend is not exhibited in the PSF subtracted PI profile, so we believe it to be spurious.  The smooth, steep incline in polarized intensity from the outer disk in to the saturation radius is the crux of the new NIR data, and explaining it is the main driver of the modeling efforts described in Section 4. 

\subsection{Intensity Images and Compatibility with the Existence of a Substellar Companion}

Direct addition of the 4 qPDI channels yields a total intensity image, which may also be diagnostic of the disk, but is dominated by stellar emission. PSF subtraction is required in order to isolate the disk component in these images. PSF subtraction is an imperfect process in systems such as SR21 for a number of reasons. First, Strehl Ratio may vary between the two sets of observations, and within each set itself, and Strehl Ratio mismatch can introduce artificial structure into the PSF-subtraction. The SR21 and HD 148212 data sets were calculated to have comparable Strehl Ratios ($\sim$20\%); therefore, we do not believe that Strehl mismatch is a dominant effect. Secondly, speckles that survive the median combination process are not necessarily cospatial in the two data sets; therefore, local excesses and deficits must be viewed with some suspicion. Thirdly, the SR21 data set was centrally saturated, while the HD 148212 data set was not, and so the counts in the SR21 image are an underestimate of the total emission. Finally, and perhaps most importantly, the total intensity of SR21 includes both a direct stellar and scattered light disk contribution, while tHD 148212's emission corresponds to direct starlight alone. These last two effects make estimation of the appropriate scaling of the PSF image for subtraction somewhat ad hoc, but by employing a series of reasonable approximations we can place bounds on the appropriate scale factor for subtraction. 

We chose to scale the PSF star HD 148212 first by the relative ratio between its 2MASS H-band magnitude and that of SR21, a quantity which we will call $a$. Assuming that the total H-band magnitude of these systems has not varied significantly with time, subtracting the PSF star intensity image scaled by this value is an oversubtraction, both because some portion of SR21's H-band magnitude corresponds to its disk contribution and because the SR21 image is saturated at the center. Therefore we use this subtraction (SR21 - $a$*HD 148212) as an ``upper limit" on the disk's total intensity in scattered light. 

We then created a grid of PSF subtractions with successively lower scale factors applied to the PSF star. We used the ratio of scattered polarized light to total scattered light in the disk as a metric in establishing the lower limit for the PSF scale factor. This ratio of polarized to total scattered light (P=PI/I$_{disk}$) should be on the order of 50\% in the case of 100\% efficient scattering of unpolarized starlight off of spherical grains; however, variables such as porosity, oblateness, and alignment as well as particle size and composition effects, multiple scattering effects, wavelength dependence and inclination effects have been invoked to explain observed deviations from this value \citep[][e.g]{Whitney:2002, Simpson:2009}.  Observed scattered polarization fractions vary from $\sim$10\% to $\sim$70\% in transitional disks, with typical values of approximately 40\% (Dean Hines and Glenn Schneider, private communication based on HST program GO 10178, October 2012). 

We deem values of $<$10\% to be unphysically low based on these results; therefore, we apply the scale factor resulting in P=10\% (0.5$a$) as our lower limit for PSF subtraction. Shown in Figure 7 is the PSF subtraction resulting from a scale factor 0.75$a$, which is halfway between the upper ($a$) and lower (0.5$a$) limits established above. The image has been smoothed by a gaussian with FWHM=6 pixels to mitigate the effects of speckles on the inferred structure of the disk. 

In estimating the ratio of polarized to total disk intensity P (PI/I$_{disk}$), we use this intermediate PSF scaling and an annulus of 15$<$$r$$<$60 pixels. This region begins just outside the radius of the PSF astigmatism and ends at approximately the outer radius of the polarized intensity disk. Within this region, we have rejected all pixels where (i) the polarized intensity flux is below noise level, (ii) P$>$1.0, indicating a PI speckle, or (iii) P$<$0, indicating an I speckle. Our upper and lower limits for the subtraction tell us that SR21's P value lies somewhere between 10 and 27\%, and a=0.75 gives P=18\%. 

The final PSF-subtracted and masked intensity images of SR21 for all scale factors investigated reveal two structures, both of which are also present in the PI data. The first is an excess in the NE quadrant near the location of the E09 companion, and the second is an arc extending upwards towards the north. These are marked as A and B in Figure 7. Although it is entirely possible that these features are spurious, they are intriguing in the context of the E09 observations and our own observation of the trend in PA with radius in the SR21 disk, which is suggestive of structure. 

E09 inferred a warm ($\sim$700 K), extended ($\sim$40 $R_\odot$)
companion 18 AU from SR21 based on the system's mid-IR flux and
visibilities.  Such a source would contribute $\sim$5\% of SR21's
bolometric flux, or $\sim$0.4 $L_\odot$, and would need to be a
low-mass star surrounded by a dusty shroud or disk.  The enshrouded
companion would likely be considerably brighter in the $H$-band than a
$\sim$700 K blackbody, itself a magnitude 16 source at SR21's
distance.  We note that a bolometric luminosity of 0.4 $L_\odot$ is
characteristic of a $\sim$0.2 $M_\odot$ T-Tauri star at 1 Myr, which
would have an effective temperature of $\sim$3000 K
\citep{Kenyon:1995}. The observed excess at point A in our data could also be due to a simple overdensity in this region of the disk.

We obtained our images of SR21 with the image rotator off, allowing us
to use Angular Differential Imaging (ADI) to search for a companion.
A traditional ADI analysis \citep{Marois:2006}
detected no statistically significant sources.  Because of SR21's
declination, we were only able to obtain $\sim$10$^\circ$ of field
rotation, making it difficult to apply more sensitive techniques like
Locally Optimized Combination of Images
\citep[LOCI,][]{Lafreniere:2007}.  We therefore
attempted a variation of LOCI using the ACORNS-ADI pipeline
\citep{Brandt:2012}, in which we supplmented the
comparison frames with 30 principal component images culled from a
large library of observations of single stars.

We detect no statistically significant companion candidates down to a
limiting contrast of 100 at $0.\!\!\prime\prime14$ (18 AU at the
distance to SR21). This is not a meaningful limit for a substellar companion.  However,
it does likely rule out a stellar companion sufficiently luminous to
account for the mid-IR excess reported by E09.  Follow-up observations
at longer wavelengths or with extreme adaptive optics will be
necessary to reach fainter detection limits.

\section{Modeling}

To assist in interpreting our results, we ran a suite of disk models using the three dimensional radiative transfer Whitney code \citep[see][for details]{Whitney:2003A,  Whitney:2003}. The key capabilities, modifications from previously reported versions and input assumptions of the Whitney code are described in Section 4.1  

In Section 4.2, we describe our efforts to reproduce the sub-mm observations of A11 using an independent code, and our verification of the A11 best fit parameters for the large grain dust disk. 

In Section 4.3, we report on several degenerate explanations for the three features that make the SED of SR21 unique: (1) the steep 12-20$\mu$m rise, (2) the marked NIR excess and (3) the very small size of the 10$\mu$m silicate feature.  

In Section 4.4 we report on our attempts to reproduce the observed steep $r$$^{-3}$ polarized intensity radial profile, and its ability to constrain the geometry and optical depth of the disk. 

Section 4.5 details our efforts to merge these three pieces into a single self-consistent model, and in Section 5 (Discussion) we speculate on the implications of our best fit model. 

\subsection{The Whitney Code and Input Assumptions}
The Whitney code mimics settling and grain growth by specifying the composition and structure of a ``large grain" disk separately from that of a ``small grain" disk. The large grain disk can be artificially settled relative to the small grain disk by setting its scale height parameters to lower values. Although both disks contain a distribution of grain sizes, we will refer to them simply as the ``large grain disk" and ``small grain disk" throughout.  

The Whitney code treats both thermal emission and scattered light in the disk. Grain temperatures are computed using the method of \citealt{Lucy:1999}. and the code treats anisotropic scattering using a Henyey-Greenstein function \citep{Henyey:1941}. The albedo, average scattering angle and maximum polarization are pre-computed at each wavelength based on a Mie calculation for the ensemble of grains in each population.  The polarization phase function is approximately dipolar, following  \citealt {White:1979}.

The Whitney code utilized herein has been modified from previously reported versions and is described in detail in \citet{Dong:2012} and Whitney et al. 2013 (in prep). The modifications of particular relevance to this paper were threefold. First, the disk density distribution was modified in order to allow for an exponential cutoff in mass as a function of radius, making it identical to the disk mass distribution of models run by A11. Secondly, the new models allow for a ``puffing up" of the inner disk edge and the inner edge of the disk gap, a physical situation widely invoked to explain NIR spectral excess \citep[e.g.,][]{Dullemond:2010} and increasingly invoked to explain the observed features of gapped disks \citep[e.g.,][]{Jang-Condell:2012}. The final modification made to the code allows for the differential depletion of large and small grains, in addition to independent geometries between the two disks (a feature included in earlier versions of the code).  This means that the disk radial density exponent ($\alpha$, $\rho$$\propto$$r$$^{-\alpha}$), radial scale height exponent ($\beta$, Z$\propto$$r$$^{-\beta}$), absolute scale height at 100AU and depletion ($\delta$) are all specified independently inside and outside of a disk gap or cavity. In effect, this allows for four independent disk components - large and small grain disks inside and outside of a specified gap radius. The size distribution and composition (and therefore opacity as a function of wavelength) of large and small grains are also specified independently.
	
The basic model input parameters (e.g., stellar mass, radius, temperature) were determined based on values in the literature, and those adopted by A11 in particular. The choices of model parameters are given in Table 2. In total, over 250 independent models of the SR21 disk were calculated in this modeling effort. 

Initial input assumptions included a disk mass of 0.006M$_{\sun}$, a gas:dust ratio of 100:1, and a large grain dust mass fraction of  85\%.  Neither of these values is well constrained by our H-band observations, so we have assumed the same values as A11. We have fixed the composition of the settled dust disk to the grain prescription of \citep{Wood:2002}, a power-law grain size distribution with an exponential cutoff \citep{Kim:1994} composed of amorphous carbon and astronomical silicates with solar abundance constraints for carbon and silicon \citep[C/H $\sim$ 3.5 x10$^{4}$ and Si/H $\sim$ 3.6 x10$^{5}$, respectively;][]{Anders:1989, Grevesse:1993}. The functional form of this size distribution is \begin{displaymath} n(a)da=C_{i}a^{-p}exp\left[-\left(\frac{a}{a_c}\right)^{q}\right]da \end{displaymath}, where $a$ is the grain size, $n$ is the number of grains, and the parameters $p$, $a_{c}$, and $q$, which control the distribution shape, are adjusted to fit the wavelength dependence of the observations. In particular, the long wavelength slope of the SED and the peak of the MIR excess are sensitive to these parameters. Following the \citealt{Wood:2002} best fit model for the HH30 disk, we use $p$=3.0, $q$=0.6 and $a_{c}$=50$\mu$m. This is a good match to the grain properties inferred by A11.

Several small grain dust prescriptions were also investigated, and the observed SED and H-band images were found to be relatively insensitive to their particulars. We use the interstellar medium dust prescription of \citealt{Kim:1994} which has a smaller average grain size than our other models. Like the large grain disk, the small grains follow a power law size distribution with an exponential cutoff, however in this case they are composed of silicates and graphite, with maximum grain sizes of 0.16$\mu$m and 0.28$\mu$m respectively. 

\subsection{Sub-mm Image Modeling}
Our H-band PDI observations probe only scattered light from the upper surface of the disk and therefore do not constrain the properties of the large grain disk where the majority of the disk mass lies. However, we have endeavored to match the sub-mm observations of A11 in addition to our newly-acquired data  in order to develop the most physically motivated disk model possible.  As a first test of the compatibility of the A11 models with the Whitney code, we began with input parameters for the disk that were identical (or nearly) to the best-fit model of A11. 

Figure 8 shows the SMA image taken by A11 in panel 1, and the equivalent 880$\mu$m image produced by our model in panel 2. We are able to produce a similarly well-fitting model with the Whitney code and identical input parameters. Model outputs were sampled with the same spatial frequencies observed with the SMA, and images were created with the same inversion and deconvolution parameters as the observations.

Having shown that the A11 and Whitney models produce equivalent results with identical input assumptions, we then investigated the effect of gap depletion on the sub-mm images to test the robustness of the A11 assumption that the gap is depleted by a factor of $\delta$=10$^{-5.9}$ relative to the outer disk. Keeping all other disk parameters fixed, we varied only the depletion factor $\delta$ in the interior of the gap by an order of magnitude for each model from $\delta$=10$^{-8}$ (two orders of magnitude higher depletion than the A11 assumption of $\sim$10$^{-6}$) to zero depletion (which, at least on the surface, is the most consistent with the evidently undepleted cavity seen in our observations). 

We break these models into three categories: ``high" depletion models ($\delta$=10$^{-8}$-10$^{-6}$), ``moderate" depletion models ($\delta$=10$^{-5}$-10$^{-3}$) and ``low" depletion models ($\delta$=10$^{-2}$-10$^{0}$) and show the real 880$\mu$m image of A11, our simulated image, residuals and visibility fits for each of these models in Figures 9-11. In each case, the model images were scaled such that the total flux was equivalent to that of the real SMA observation, and the required scaling is given in the model panel of the figures. 

The ``low" depletion models provide poor fits to both the 880$\mu$m image and the visibility profile. We conclude, as did A11, that the SR21 disk is depleted in large sub-mm emitting grains interior to 36AU. All of the moderate and high depletion models provide reasonable fits to both the 880$\mu$m image and the visibility profile; however, the $\delta$=10$^{-5}$ model shows the lowest residuals and the $\delta$=10$^{-6}$ model provides the best fit to the visibility profile.  The $\delta$=10$^{-6}$ model requires just a 2\% increase in flux to sum to the observed flux density (easily explained by small differences in input opacity assumptions), while the $\delta$=10$^{-5}$ model requires an 8\% scaling, so we adopt the $\delta$=10$^{-6}$ value for the large grain $r$$<$36AU disk in all further modeling.   This value is closest to the A11 best fit value of $\delta$=10$^{-5.9}$; therefore, we consider our large grain models to be functionally equivalent. 

As a test of the robustness of the large grain disk assumptions, we verified that the small grain disk parameters specified in our models did not create observable variation in the simulated 880$\mu$m image. As expected, we found that wide variation in input small grain parameters has no observable effect on the 880$\mu$m images produced by our models. The 880$\mu$m emission of the large grain disk is therefore strongly decoupled from the small grain disk parameters.

Having verified that we can reproduce the A11 result, that the A11 inferred gap radius and depletion factor for large sub-mm emitting grains is a good match to the models produced by both codes, and that this result is unaffected by variation in small grain parameters, we treat the A11 large grain disk parameters (with the exception of disk inclination, for which we adopt our measurement) given in Table 2 as fixed throughout the remainder of our modeling efforts. 

\subsection{SED Modeling}	
There are sufficient data in the literature to construct a well-sampled observed SED for SR21. In addition to 2MASS J, H and K photometry, SR21 photometry has been obtained with Spitzer's Infrared Array Camera (IRAC) and Multiband Imaging Photometer (MIPS) at 3.6, 4.5, 5.8, 8 and 70$\mu$m, the AKARI spacecraft at 18$\mu$m, the Infrared Astronomical Satellite (IRAS) at 12, 25 and 60$\mu$m, the Submillimeter Array (SMA) at 880$\mu$m, and the James Clerk Maxwell Telescope's Submillimetre Common-User Bolometer Array (SCUBA) at 350, 450, 850 and 1300$\mu$m. 10-35$\mu$m spectra from the Spitzer Infrared Spectrograph (IRS) and 52-97$\mu$m from the MIPS SED mode are also available.  These are shown overplotted on all SEDs generated by our models. Although we show it for completeness, the MIPS 70$\mu$m photometry point has been excluded from our fits due to its inconsistency with both the MIPS SED spectrum and IRAS 60$\mu$m photometry. All reported photometric points and spectra have been extincted according to a standard $r$$_{V}$=3.1 interstellar extinction law \citep{Cardelli:1989}. The extinction for SR21 has been estimated to be as high as A$_{V}$=9.0 \citep{Prato:2003}, but A11 have argued that the published photometry is most consistent with a lower value of A$_{V}$$\sim$6.3. We also find that lower extinctions are most consistent with the observed optical/NIR SED, so we adopt the same A$_{V}$=6.3 value for SR21.

The SR21 SED shows circumstellar dust excess emission longward of the stellar blackbody peak, as evidenced by the long wavelength excess (see Figure 12). Unlike protoplanetary systems, which tend to have flatter transitions between the stellar and dust peaks in the SED indicating significant dust populations at all temperatures, the SR21 SED has a distinct double peaked profile, reaching a local minimum near 12$\mu$m. This suggests that there is a dearth of hot grains in the disk, a conclusion supported by the observed sub-mm cavity. Debris disk systems, on the other hand, are much more depleted of small grains, and their emission traces the stellar photosphere out to wavelengths of 20$\mu$m or more. In contrast, the SR21 SED rises rapidly beyond 12$\mu$m, indicating a healthy population of warm grains that emit at $\lambda$$>$12$\mu$m.  

Shortward of the 12$\mu$m local minimum in the SR21 SED, the emission is also greater than photospheric, which indicates a small population of very warm grains (a hot inner disk component) or, alternatively, a warm companion. The SR21 SED shows little evidence of a 10$\mu$m silicate feature, while many similar transitional disk systems exhibit strong, broad silicate emission at this wavelength. A good disk model for SR21 therefore needs to reproduce all three of these distinct SED features: (a) the steep 12-20$\mu$m rise, (b) the sub-12$\mu$m excess and (c) the lack of a strong, broad 10$\mu$m silicate feature.  

The Whitney code produces stable SEDs with relatively few ($\sim$10$^{6}$) photons, making it much less computationally expensive than generating images/radial profiles. We were therefore able to explore much of the available parameter space for SR21 to find several classes of reasonable models capable of reproducing the observed features of the SED. 

\subsubsection{NIR Excess Fits}
The NIR excess shortward of 10$\mu$m is a distinct feature in the observed SED. The (small) excess in this region was originally attributed to a 0.07-0.25AU warm inner disk component. Indeed, adding this component to our models produces a well-fitting NIR excess. However, the MIR observations of \citet{Pontoppidan:2008} are somewhat inconsistent with the presence of this interior hot disk component. Their data show a sharp truncation in gaseous CO at 7AU, suggesting a high flux of UV photons. This feature would be difficult to produce with an inner hot disk component at $r$$<$7AU capable of intercepting them. Furthermore, allowing the large grains to fill the cavity in to 0.07AU in our models, even while depleting them by a factor of 10$^{-6}$, results in point-like sub-mm emission at the disk center, which is not present in the A11 observations. 

The presence of the warm companion predicted by E09 is another potential mechanism for creating excess NIR flux in SR21, and in this case allows for a larger inner small grain disk truncation radius. Our observations can neither confirm nor exclude the presence of a companion, yet they do provide indirect support of its existence, as detailed in Section 3.3. 

Otherwise identical models with an (a) 7AU truncation and a warm (T=730K, R=30$r$$_{\sun}$, E09) blackbody companion (b) a disk that extends inward to the sublimation radius at 0.07AU are shown in Figure 12 to demonstrate that both assumptions reproduce the observed NIR excess. Without any means of probing disk structures interior to the saturation radius of 0$\farcs$1 in our observations, we cannot break the degeneracy between inner disk extent and the existence of a companion in the NIR SED fits using scattered light imagery.

We hold a slight preference for the warm companion hypothesis based on the P08 CO results and our own observations of excess intensity and deviation from polarization axisymmetry (indirectly supporting the companion hypothesis and described in detail in Section 3.3), so we assume a 7AU truncation in all of our scattered light models. We add the warm companion component to all of our modeled SEDs in order to reproduce the observed NIR excess, but acknowledge that a significant inner disk extent is also a plausible mechanism for creating this feature. We do not add polarized intensity emission from the companion into our simulated images because the polarized intensity contribution of such a companion is (a) not well constrained by models and (b) should not have a significant effect on the azimuthally-averaged radial profile. 

\subsubsection{10$\mu$m Silicate Fits}
The final feature of note in the SR21 SED is the lack of a prominent 10$\mu$m amorphous silicate feature, which tends to be rather strong and broad in Herbig disks and is believed to originate from the surface layers of the inner (<20AU) disk \citep{van-Boekel:2005, Meeus:2001}. The IRS spectrum spans this region of the SED and also that of the weaker 18$\mu$m silicate feature without any evidence of a strong line flux at either wavelength. 

We find that the majority of the 10$\mu$m silicate emission comes from the inner disk wall in our models, and we can suppress it by either restricting the scale height of the small grain inner disk wall or by restricting small grains to a very small proportion of the total mass ($<$1\%). Generally speaking, many otherwise well-fitting models overproduce 10$\mu$m silicate emission.  Changing the small grain prescription may be warranted in future modeling attempts. Specifically, the silicate features could be suppressed by either increasing the grain sizes or limiting the proportion of silicate grains in the disk. We leave this for future modeling studies. 

Our scattered light imagery cannot give any information about the size of the small grain disk wall where the 10$\mu$m feature originates,  whether it lies at 7AU or 0.07AU (the sublimation radius). We limit the size of the inner disk wall to small scale heights in all of our scattered light models in order to suppress the 10$\mu$m feature, but in some cases not severely enough, resulting in overproduction of 10$\mu$m emission in the SEDs. We could presumably correct this by running additional models with smaller inner wall scale heights or reduced small grain populations, but have elected not to focus on reproducing this feature precisely for models that are poor fits to the radial polarized intensity profile due to the computational expense. The quality of fit to the 10$\mu$m feature was taken into account in choosing the final best-fitting disk model, described in section 4.4 and shown in figures 15-16.

\subsubsection{Fits to 10-20$\mu$m rise}

The steepness of the rise from 12-25$\mu$m is perhaps the most unique and significant feature of the SR21 SED and is the primary feature that led to its ``transitional disk" classification. The presence of this feature requires that there be a sharp increase in disk emitting area (e.g. a ``wall") at intermediate (20-40AU) radii. However, the precise geometry of the ``wall", as well as its composition, is not well constrained by the observed SED.

The simplest way to create a ``wall" is to allow for depletion inside of the $r$=36AU gap, so that the disk makes a transition from optically thin to optically thick at this radius. We know that the the disk is optically thin at $r$$<$36AU for long wavelengths (and therefore in large, thermally-emitting grains), and optically thick beyond; therefore, the large grains certainly do present a ``wall" at 36AU. All of the models that we consider are therefore ``gapped" in large grains. 

However, the sub-mm data places no such constraint on small grain depletion. In fact, there is reason to believe that small grain dust filtration occurs at gap radii \citep{Zhu:2012, Rice:2006}, and that material must survive into observed sub-mm dust gaps, as a number of gapped disks are still observed to be actively accreting \citep{Najita:2007, Espaillat:2012, Salyk:2009}. The observed H-band radial profile of the SR21 disk would seem to support this hypothesis, as it shows no evidence of a change in small grain disk properties at $r$=36AU. Until our Subaru H-band data were obtained, there were no spatially resolved data at shorter wavelengths to constrain the properties of the small grains and the amount to which they are depleted inside of the sub-mm cavity.

The simplest hypothesis, put forward by previous SED modelers \citep[A11, E09]{Brown:2007} is that the small grain disk follows the same structure as the large grain disk, including the $r$$<$36AU gap. This hypothesis reproduces the observed SED, but it is inconsistent with our new Subaru H-band polarized intensity profile. Simulated radial profiles and SEDs for various small grain depletion factors are shown in Figure 13. They show that although small grain depletion in the $r$$<$36AU region reproduces the 12-20$\mu$m rise in the SED, it is strongly inconsistent with the observed H-band radial profile. All models in which the inner disk is strongly depleted, and therefore optically thin, suggest that we should have seen an excess in H-band emission at the cavity wall and a turnover in the radial profile at this radius. 

Undepleted and low depletion models, on the other hand, are inconsistent with the observed SED. Additionally, their radial profiles are too shallow to be consistent with the H-band data. Appendix A details our efforts to find a scenario in which the small grain disk, whether optically thin or optically thick at $r$$<$36AU, contributes to the 12-20$\mu$m rise in the SED.

In the end we find that the steep 12-20$\mu$m rise can be created with a disk depleted only in large grains in its interior. Our data suggest that the small grain disk is not ``gapped". 

\subsection{Scattered Light Imagery}

At first glance, the H-band radial polarized intensity profile is difficult to reconcile with the large, heavily depleted sub-mm cavity, as it spans the region of the sub-mm wall without any evident break or turnover. In fact, the principal features of the observed radial polarized intensity profile, namely its smoothness and its steepness ($\sim$$r$$^{-3}$), were not reproducible under any of the scenarios we investigated in which the small grain disk contained a depleted cavity. Furthermore, we were unable to reproduce the radial profile under any scenario in which the H-band scattered light originates from the surface of an optically thick small grain component, whether gapped or not. This effort is described in detail in Appendix A, and representative models for each variety of optically thick disk model investigated are shown in Figures 17-19. This is a surprising result, as we had no reason to expect that transitional disks are optically thin at short wavelengths.

We find that the steepness of the radial profile requires that we invoke an optically thin small grain component that spans the entire disk. In order to create this, we must either (a) relax the A11 assumption that the small grains compose 15\% of the disk's dust mass and restrict their contribution to $<$1\%, making the small grain dust disk optically thin throughout, or (b) invoke a third optically thin disk atmosphere or envelope component that has a larger vertical extent than either the large or small grain disks.  

Under scenario (a) the optically thick ``wall" producing the 12-20$\mu$m SED rise must be composed entirely of large grains in order to allow the small grain disk to be optically thin. As we noted in section 4.3, we do not require a small grain wall in order to reproduce the 12-20$\mu$m rise in the SED. In fact, we can reproduce the SED structure at $\lambda$$>$$12\mu$m with a large grain gap alone. We do need to increase the scale height of the gap wall in order to provide enough flux at these wavelengths (relative to models where an optically thick small grain component contributes to the rise), but not by such a degree as to lead us to declare these models unphysical (Z$_{wall}\gtrsim$1AU). 

Since an optically thin small grain disk means that we inevitably ``see" the large grains at the midplane, these models require that the large grain disk not scatter efficiently at H-band. Otherwise, they suffer from the same local maximum problem that led us to exclude ``Gapped" small grain disks.  As can be seen in Figure 14, which shows a scenario under which the small grain disk is optically thin throughout, the \citet{Wood:2002} dust prescription chosen for the large grain disk (and found to be a good match to that of A11) does scatter weakly at H-band, resulting again in a turnover in the radial profile at 36AU. These models would require an alternate grain size distribution and/or composition in order to prevent a local excess at 36AU. We have not endeavored to explore this possiblity here.  

We disfavor explanation (a) because it requires that small grains be optically thin throughout the vertical extent of the disk. An entirely optically think small grain disk disk is very unlikely given the small amount of disk mass allowed and the observed optically thick $r$$>$36AU large grain component. In order to be optically thin throughout, we require not only that $<$0.1\% of the disk mass lie in the small grain disk, representing a likely unphysical degree of grain growth or small grain depletion, but also that the large grain disk's grain size distribution contain no grains that scatter at H-band, an equally unlikely scenario.  

Under scenario (b), which we believe to be far more likely, an envelope component lies above the optically thick small grain dust disk with a significant vertical extent. This envelope contains very little total mass (on the order of 10$^{-7}$M$_{\sun}$) and is composed of small ISM-like grains \citep{Kim:1994} Representative model outputs for this scenario are shown in Figure 15, and a schematic of the disk structure is shown in Figure 16. 

Although we call this third disk component an ``envelope", it is not clear that it fits into the canonical definition. There has been no evidence in the literature to suggest that SR21 is still actively accreting, and the only constraints on the accretion rate are upper limits \citep{Natta:2006}. This is in contrast to a number of other gapped disks that do show evidence of accretion \citep{Salyk:2009}. 

It seems that this optically thin component may be more readily explained as a vertically extended disk ``atmosphere" of sorts, rather than an accretion envelope. Whether this is part of the natural structure of the disk, a remnant of the envelope, or a disk wind launched from the optically thick disk surface cannot be determined from our data. In fact, our observations are relatively insensitive to its precise geometry. The radial profile of the optically thin component follows an $r$$^{-3}$ power law in all cases where the disk $\alpha$-$\beta$=-1, which allows for a large range in flaring and density structures. 

Under this three component disk model, the radial profile is still sensitive to the structure of the underlying optically thick small grain disk surface. In particular, the addition of the optically thin component that gives rise to the $r$$^{-3}$ power law cannot mask a discontinuous small grain distribution at the midplane, even when the optically thin component is allowed to contribute an order of magnitude more scattered light than the optically thick component. A large number of optically thick disk models were investigated in this effort, and are described in detail in Appendix A. 

\section{Modeling Discussion}
Our requirements to fit the radial profile are threefold. First, we require that there be enough small grain material inside the 36AU large grain cavity to scatter direct NIR starlight before it reaches the cavity wall, effectively shadowing it. Although this gap must be optically thin to sub-mm light in order to reproduce the SMA observations of A11, we find no viable scenarios in which it can be optically thin at shorter wavelengths and remain consistent with the observed H-band radial polarized intensity profile. 

Secondly, the smoothness of the observed H-band polarized intensity profile suggests that the surface of this underlying optically thick small grain disk must follow a smooth surface geometry. We cannot fit our data under scenarios where there is a marked effect, either in differential depletion or a scale height discontinuity, of the large grain dust gap on the small grain distribution. This is perhaps the most interesting and puzzling result of our observations and analysis -- that the small H-band scattering grains in the SR21 disk must follow a smooth distribution that is relatively unaffected by the goings-on at the disk midplane. This supports scenarios under which dust filtration occurs at the large grain gap wall, allowing the small grains to be carried into the gap along with the disk's gas. 

Finally, we find that the steepness of the H-band radial profile of SR21 is indicative of an optically thin small grain disk component. In order to remain consistent with the first two criteria, this requires that we invoke a third vertically extended optically thin disk component overlying the optically thick portion of the disk. 

As noted in Section 2,  the AO system performance was sub-optimal during the SR21 observations, and uncorrected effects such as those of a polarized seeing halo may mimic disk emission, particularly close to the star. Thus it is possible that the true SR21 radial profile flattens out in the interior and we are mistaking uncorrected PSF effects for disk emission in the inner region. However, all pixels in the saturated region were excluded from the radial profiles, and the PSF artifacts do not extend beyond the innermost bin in our radial profile plots, so this is unlikely. 

The computational expense of scattered light models means that we did not investigate the scattered light parameter space as systematically as the SED parameter space, and a more thorough and systematic exploration of the available parameter space is warranted to ensure that our solution is unique. 

It could be argued, however, that ADI observations with more field rotation, higher AO performance (Strehl Ratio) H-band observations and/or further multiwavelength data that resolve the mysterious SR21 ``gap" for other varieties of disk grain are equally warranted. Observing methods that are sensitive to sharp disk features (e.g., ADI) could help verify whether there is any local excess in scattered light at the 36AU dust wall in addition to resolving the potential companion. 

It is also worth noting that all of our disk models have assumed axisymmetry. Unresolved spiral structure, emission from a companion, disk warps, etc. could be present in the SR21 disk and may have an effect on the radial profile that has not been examined in this analysis. 

Uniformity of grain composition as a function of radius is another aspect of the models that is perhaps unphysical. Compositional differences, such as higher volatile abundances in the gapped region, could allow for a more marked effect of density on surface brightness in the inner disk, allowing optically thick large-$\beta$ large-$\alpha$ models to increase more steeply towards the star. 

\section{Conclusion}

We have spatially resolved the SR21 transitional disk for the first time at NIR wavelengths. Our data show the signature of an optically thin small grain disk component, and place strong constraints on the underlying distribution of any optically thick surface grains. 

Our Subaru data resolve the disk in scattered polarized light for stellocentric $r$$\geqslant$0$\farcs$1 ($\gtrsim$12 AU). Extended polarized emission is present above noise level out to $r$$\sim$0$\farcs$6 ($\sim$ 80 AU) and exhibits a centrosymmetric pattern of polarization vectors.  The azimuthally-averaged radial profile of the polarized intensity emission reveals that the disk flux drops off steeply ($r$$^{-3}$) and smoothly with radius. The disk is also evident in the PSF subtracted total intensity image, and the range of viable scale factors for PSF subtraction give us an estimated polarization percentage for the disk of 10\%$\lesssim$P$\lesssim$27\%. 

We confirm with an independent model that scenarios in which large midplane grains are depleted by a factor of $\sim$10$^{-6}$ inside of a $\sim$36 AU gap reproduce the SMA observations of A11; however, small grains cannot be depleted to the same degree inside the cavity.

Our analysis eliminates a variety of scenarios under which there is a discontinuity in the small grain disk at the 36AU large grain gap radius, both in terms of physical likelihood and ability to fit the observational data. In order to be consistent with the lack of an equivalent gap in the Subaru HiCIAO H-band polarized intensity images and radial intensity profiles, there must be a decoupling of large and small grain disk parameters in the SR21 disk. Our data rule out scenarios in which the SR21 disk contains an optically thin small grain dust gap with a wall at 36AU. 

Although we can neither confirm nor deny its existence definitively, we present several pieces of anecdotal evidence in favor of a companion in the SR21 disk. These include an intensity excess near the appropriate location of E09's  hypothesized companion, a deviation from centrosymmetry in the polarization vectors in this region, and an apparent migration of the disk major axis from inner to outer disk, suggesting a disk warp or potentially unresolved spiral structure.  

We conclude that the best fitting models for the H-band radial polarized intensity profile, the observed sub-mm emission and the multiwavelength SED for the SR21 disk invoke three separate components. These axisymmetric, uniform radial grain composition scenarios suggest an optically thin envelope or atmosphere component lying above an optically thick, geometrically smooth small grain disk component. Further constraints on the geometry of this very interesting transitional disk may be obtained by follow-up at higher spatial resolutions and at other wavelengths. 

\begin{acknowledgements}
We gratefully acknowledge funding from the National Science Foundation East Asian and Pacific Summer Institute Fellowship (Follette), NSF-AST 1009314 (Wisniewski) and NSF-AST 1009203 (Carson). We are grateful to Collette Salyk, Glenn Schneider, Dean Hines, Don McCarthy, Vanessa Bailey and Johanna Teske for their insightful comments. The authors wish to recognize and acknowledge the very significant cultural role and reverence that the summit of Mauna Kea has always had within the indigenous Hawaiian community. We are most fortunate to have the opportunity to conduct observations from this mountain. Part of this work was carried out at JPL. This work is partly supported by a Grant-in-Aid for Science Research in a Priority Area from MEXT.
\end{acknowledgements}

\clearpage

\begin{table} [ht]
\caption{Position Angle Estimates}
\centering
\begin{tabular} {c c c}
\hline\hline
PA & Radius probed & Reference\\
$^{\circ}$ & AU & \\
\hline
15$\pm$4 & 7.0$\pm$0.4 & P08 (v=1-0) \\
16$\pm$3 & 7.6 $\pm$0.4 & P08 (v=2-1) \\
44$\pm$7 & 11$\pm$1 & E09 \\
61$\pm$8 & 15$\pm$1 & E09 \\
76$\pm$8 & 12-25 & this work \\
96$\pm$7 & 25-65 & this work \\
100 & 30-120 & A11 \\
105 & 30-120 & Brown 2009 \\
\hline
\end{tabular}
\label{table:inputs}
\end{table}

\begin{table} [ht]
\title{Model Input Parameters}
\centering
\begin{tabular} {c c}
\hline\hline
Parameter & Model Input(s)\\
\hline
\multicolumn{2}{c}{Input Assumptions}\\
M$_{star}$ & 2M$_{\sun}$$^{*}$\\
R$_{star}$ & 3.15R$_{\sun}$$^{*}$\\
T$_{star}$ & 5750K$^{+}$\\
A$_{V}$ & 6.3\\
M$_{disk}$ & 0.006M$_{\sun}$$^{*}$\\
$r$$_{gap}$ & none, 18, 36$^{*}$ AU\\
$r$$_{disk, in}$ & $r$$_{sub}$-7AU\\
$r$$_{disk,out}$ & 200AU\\
$i$ & 14, 22$^{*}$\\
$r$$_{C}$ & 15AU$^{*}$\\
\hline
\multicolumn{2}{c}{Large Grain Disk Parameters}\\
composition & \citealt{Wood:2002}$^{+}$\\
f$_{d}$ & 0.85$^{*}$\\
Z$_{100}$ & 1.52AU$^{*}$ \\
$\alpha$$_{outer}$ & 2.15$^{+}$\\
$\beta$$_{outer}$ & 1.15$^{+}$\\
$\alpha$$_{gap}$ & 2.15$^{+}$\\
$\beta$$_{gap}$ & 1.15$^{+}$\\
$\delta$ & 10$^{-8}$,10$^{-7}$,10$^{-6^{+}}$,10$^{-5}$, 10$^{-4}$, 10$^{-3}$, 10$^{-2}$, 10$^{-1}$, 1\\
\hline
\multicolumn{2}{c}{Small Grain Disk Parameters}\\
composition & \citealt{Kim:1994, Cotera:2001}\\
f$_{d}$ & 0.15\\
Z$_{100}$ & 2-15AU \\
$\alpha$$_{outer}$ & 0.5-4\\
$\beta$$_{outer}$ & 1-3.15\\
$\alpha$$_{gap}$ & -5-10\\
$\beta$$_{gap}$ & 1-3.15\\
$\delta$ & 10$^{-8}$,10$^{-7}$,10$^{-6}$,10$^{-5}$, 10$^{-4}$, 10$^{-3}$, 10$^{-2}$, 10$^{-1}$, 1, 2, 5, 10\\
\hline
\end{tabular}
\label{table:inputs}
\caption{The range of Whitney model input parameters investigated for SR21. Where several discreet values were attempted, the table entry is given in the form of a list, and for parameters where a range of values was attempted, the table entry gives the full range of attempted values. Parameters identical to those of A11 are marked with an asterisk, while equivalent parameters (rounded for convenience, compatibility, or translated to Whitney code syntax) are marked with a + sign. R$_{C}$ is the characteristic radius for the exponential density cutoff as defined in A11, f$_{d}$ is the fraction of the disk's dust mass, Z$_{100}$ is the disk scale height at 100 AU, $\alpha$ is the exponent for the radial midplane density dependence, $\beta$ is the exponent for the radial scale height dependence, and $\delta$ is the depletion factor of the gapped region relative to the outer disk}
\end{table}

\clearpage

\begin{figure}
\includegraphics[scale=.8]{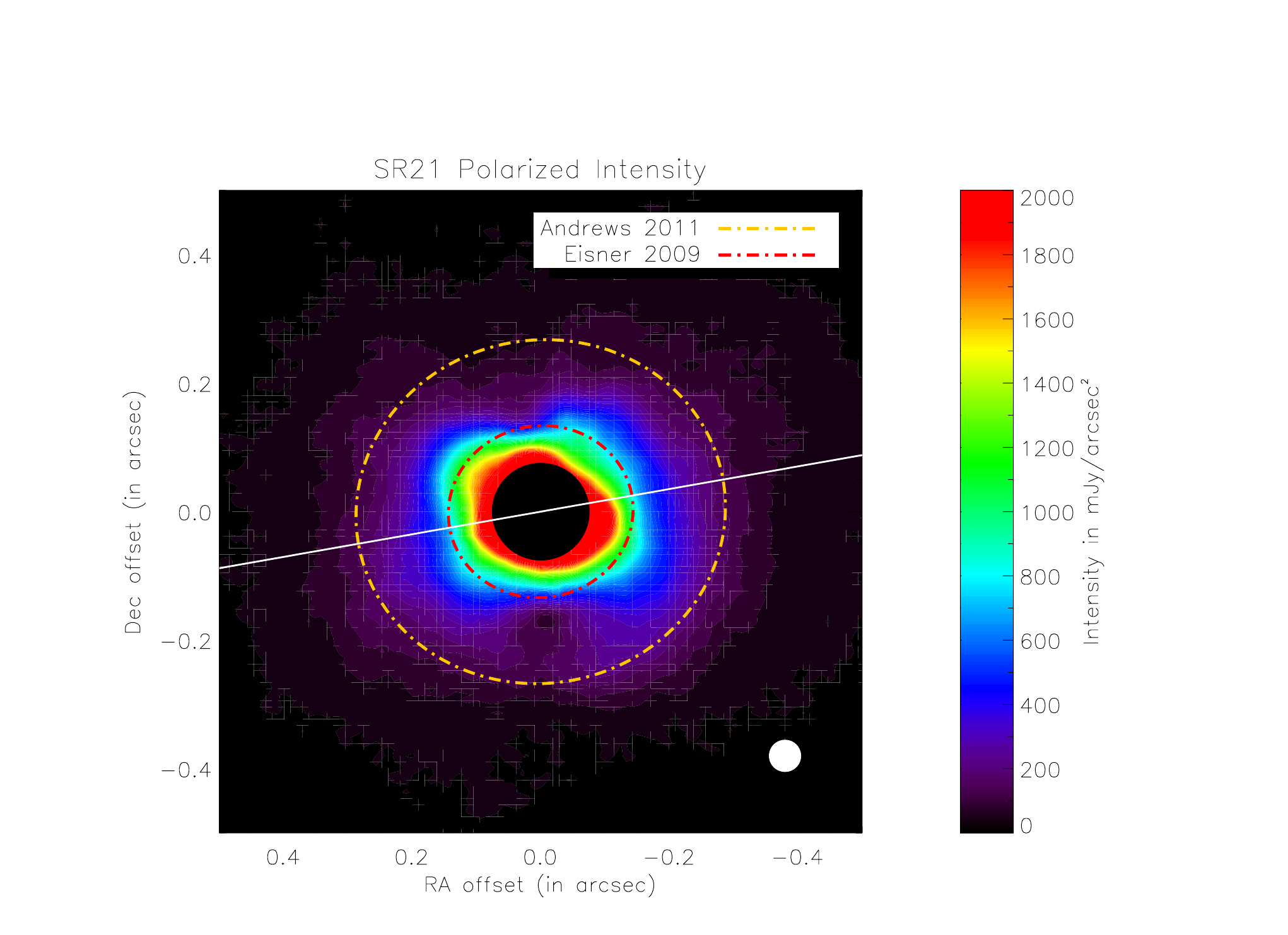}
\caption{Median polarized intensity (PI) image of SR21 at H-band. The gap radius inferred by A11 and the companion orbit inferred by E09 are indicated with dashed ellipses. These are circles inclined at 14$^{\circ}$ and rotated to a major axis PA of 100$^{\circ}$ (A11) and 60$^{\circ}$ (E09) (indicated with the solid white and red lines). North is up; east is left. The central 0$\farcs$07 were saturated in the raw images and have been masked out. A diffraction limited spot is shown at the bottom right for reference. Modeling suggests that the emission deficit $\sim$0$\farcs$2 south of the central source is a result of the disk geometry and is not a physical structure. \label{fig1}}
\end{figure}

\begin{figure}
\includegraphics[scale=.8]{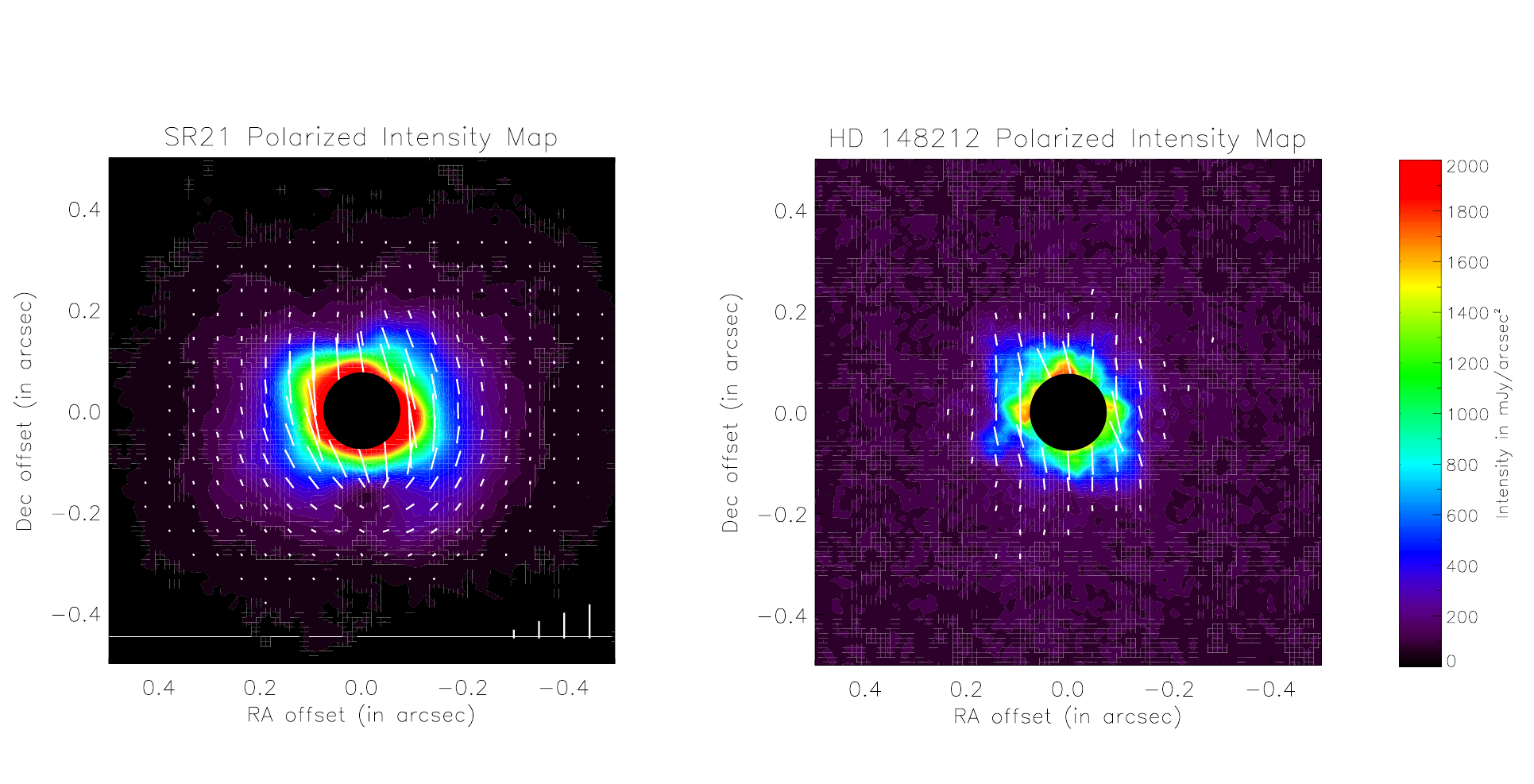}
\caption{Left: Polarization vectors overplotted on the median polarized intensity (PI) map of SR21. A general centrosymmetric geometry is evident in the SR21 vectors, suggesting that this emission does indeed trace scattered light from a circumstellar disk. Right: Polarization vectors for the PSF star HD 148212 overplotted on its median PI image, which has been scaled by a factor of 3 to compensate for the difference in exposure time. In this case, the vectors do not show a coherent pattern, and PI emission is weak and compact. This is interpreted as photon noise residuals and is used to estimate the error in the polarized intensity profiles shown in figure 3. In both cases, vectors are scaled by the strength of the polarized intensity signal in the region and are not plotted when they fall at or below noise level. Vectors whose lengths correspond to 500, 1000, 1500 and 2000mJy/arcsec$^{2}$ are shown in the bottom right of the SR21 panel as a scale reference.\label{fig2} }
\end{figure}

\begin{figure}
\includegraphics[scale=.8]{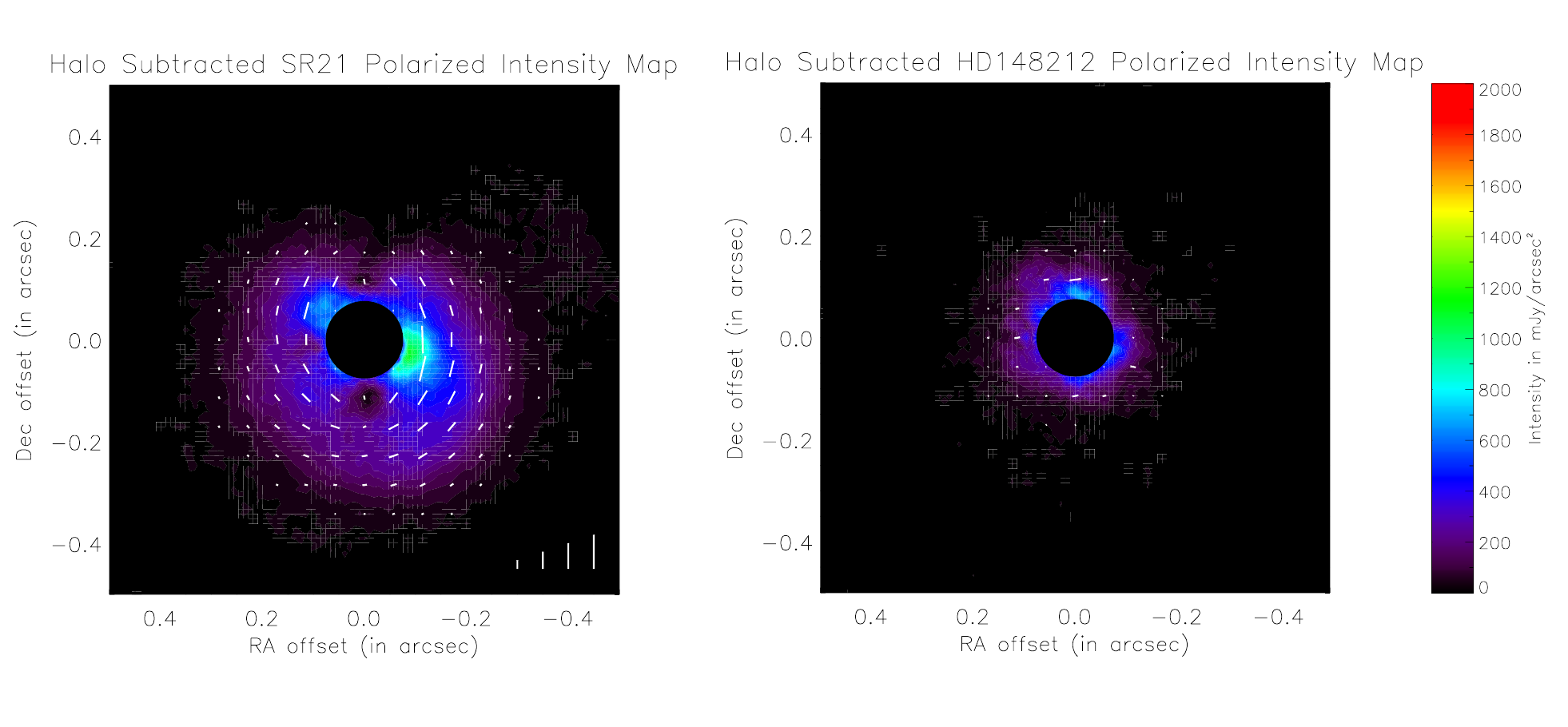}
\caption{SR21 polarized intensity image (background) and polarization vectors (overplotted) after first order correction for a seeing limited halo component. This method reveals the polarization vectors to be more centrosymmetric, particularly along the disk minor axis) than originally inferred. Local deviations from centrosymmetry persist in the northeast and southwest quadrants of the disk, where unresolved disk structures and/or companions could induce local polarization structure that deviates from the larger centrosymmetric pattern. The process of subtracting this halo component has introduced additional PSF artifacts into the background PI map, most evident in the negative astigmatic pattern centered on the star. We have thus chosen to model the uncorrected polarized intensity map, which we believe more closely resembles the true structure of the SR21 disk. Vectors whose lengths correspond to 500, 1000, 1500 and 2000mJy/arcsec$^{2}$ are shown in the bottom right of the SR21 panel as a scale reference.}\label{fig3} 
\end{figure}

\begin{figure}
\includegraphics[scale=0.8]{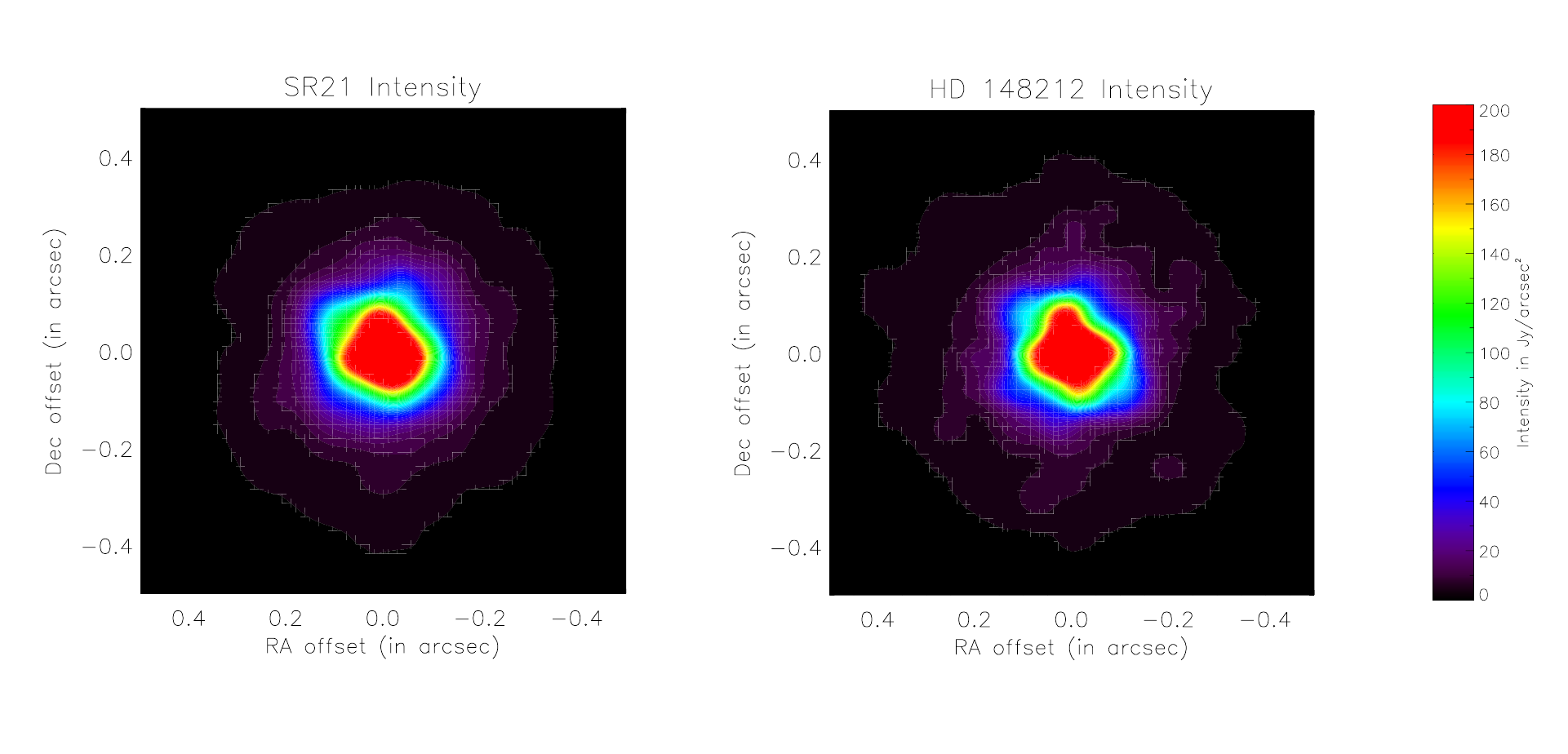}
\caption{Intensity images of SR21(left) and HD 148212 (right). HD 148212 has been scaled according to differences in exposure time and intrinsic H-band magnitude (based on published 2MASS values) to match SR21 for subtraction. Some uncorrected astigmatism is evident in the four lobes of the PSF in both stars.}\label{fig4} 
\end{figure}

\begin{figure}
\includegraphics[scale=1]{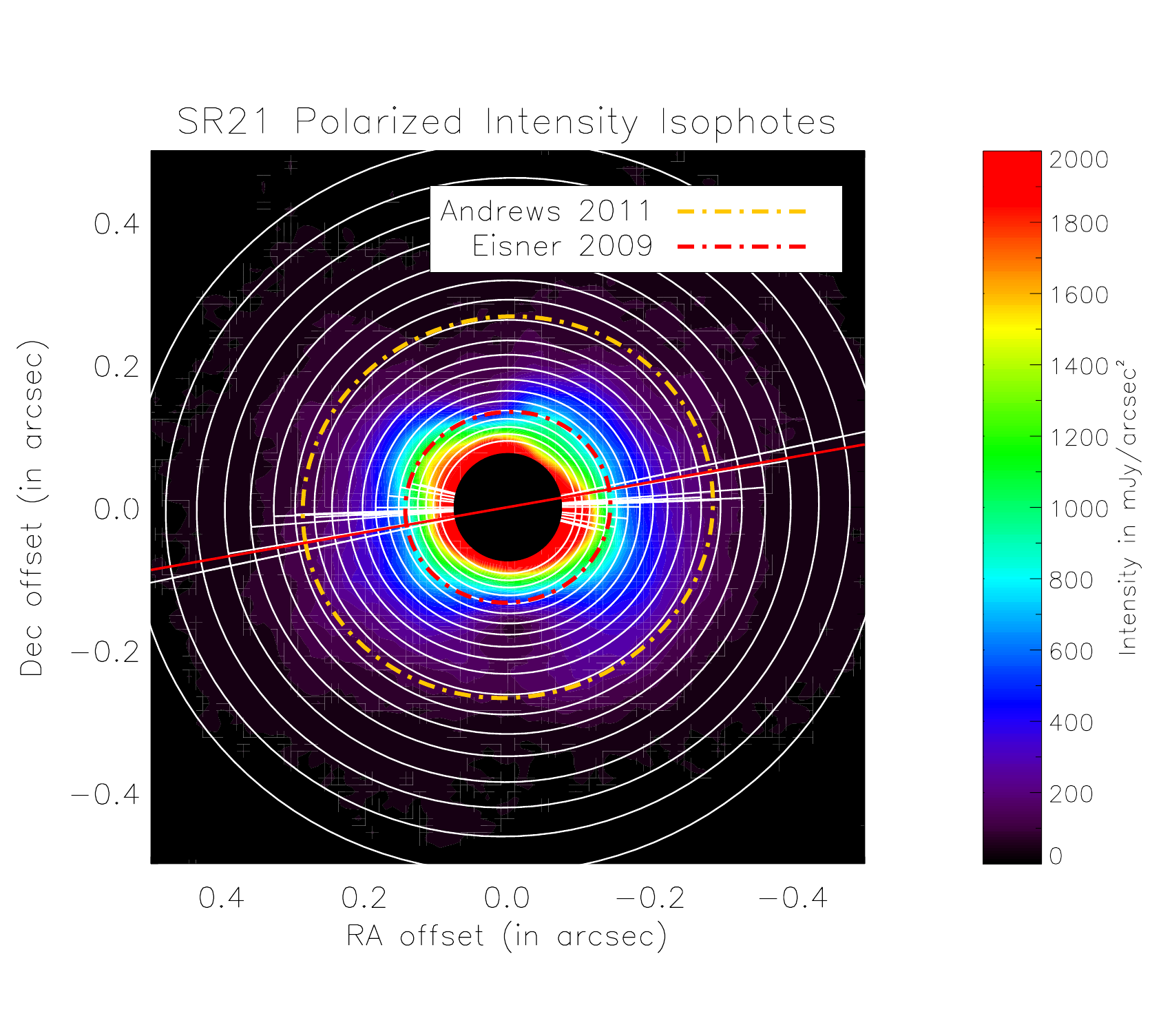}
\caption{Elliptical isophotal fits to the SR21 Polarized Intensity image. The fits are shown overplotted on the PI image in white starting at a semimajor axis of 10 pixels (well outside the saturation radius) to a semimajor axis of 50 pixels (where the PI flux drops to noise level). The major axis of each isophote is overplotted as a straight white line. The average value of the major axis position angle is clocked by $\sim$15$^{\circ}$ from the inner to outer disk, suggesting unresolved disk structure. The ellipticity of the isophotes was also used to estimate the inclination value of 14$^{\circ}$ (where 0$^{\circ}$ would be face-on).}\label{fig5} 
\end{figure}

\begin{figure}
\includegraphics[scale=.9]{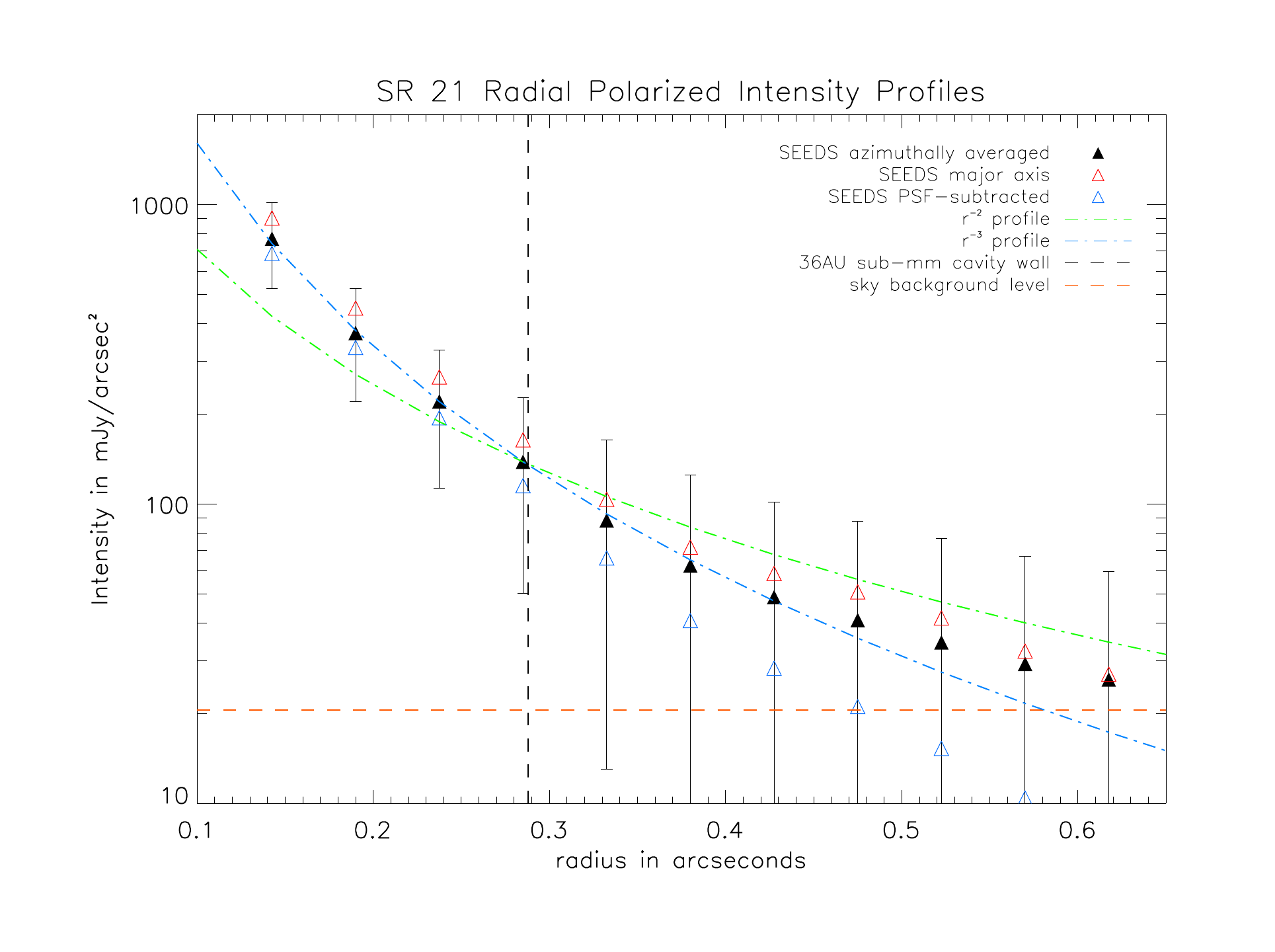}
\caption{Radial polarized intensity profiles for SR21 binned to the H-band diffraction limit (0$\farcs$5) and shown as triangles. The black filled triangles represent the azimuthally-averaged radial profile and the red open triangles represent the profile as measured in an 11 pixel wide region centered on the major axis of the disk. The blue open triangles represent the profile of the PSF-subtracted image. All three profiles follow a very similar slope, with the major axis profile systematically higher in flux per pixel by ~20\% relative to the azimuthally-averaged profile, and the PSF subtracted image systematically lower, as is to be expected. Because the images were saturated interior to 0$\farcs$07, radial profiles are only shown for bins that fall completely outside of the saturated region (from 0$\farcs$1 outward). Also shown is a canonical $r$$^{-2}$ profile in green and a $r$$^{-3}$ profile in blue. The $r$$^{-3}$ profile is a better fit to the data, although there is some evidence that it becomes shallower in the outer disk beyond 0$\farcs$45. The veracity of this shallower profile in the outer disk is brought into question by the steepness of the PSF subtracted profile, suggesting that the flattening in the outer profile may be due to halo effects. Errors were estimated from the photon noise in the four individual channels used to create the PI image summed in quadrature (black error bars). The location of the sub-mm cavity wall at 36AU and the noise floor are indicated with vertical and horizontal dashed lines, respectively.} \label{ fig6}
\end{figure}

\begin{figure}
\includegraphics[scale=0.8]{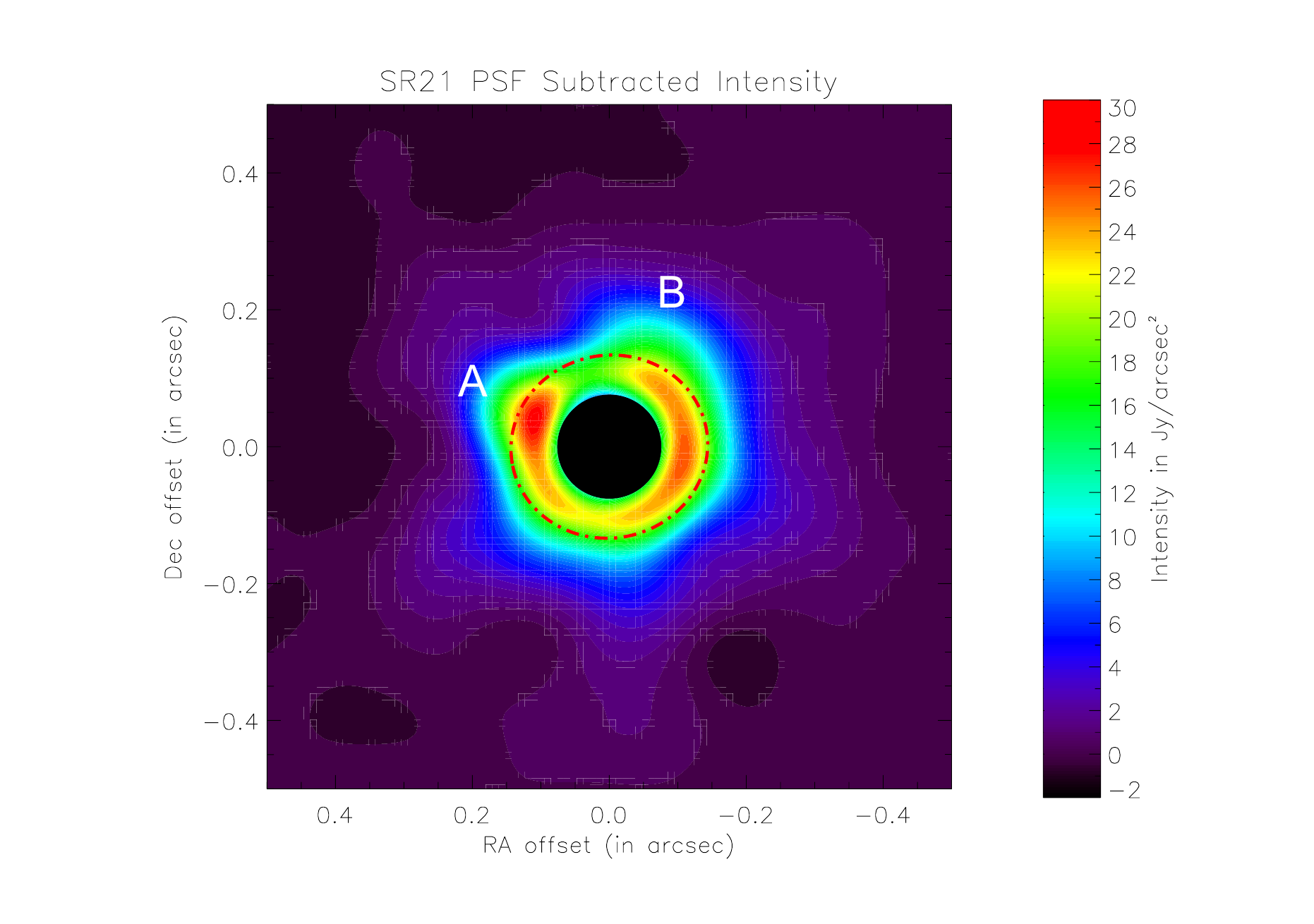}
\caption{H-band PSF-subtracted intensity image with the 18AU orbit of the companion proposed by E09 overplotted. There is a distinct excess of emission in the Northeast quadrant (A) at a location that is consistent with their hypothesized companion, although its proximity to PSF artifacts makes robust identification impossible with this data. There is also a northward extension of the disk that is nominally present in the PI image (B). The image has been smoothed with a FWHM=6 pixel Gaussian in order to suppress artificial structures induced by speckle noise.}\label{fig7} 
\end{figure}

\clearpage

\begin{figure}
\includegraphics[scale=1]{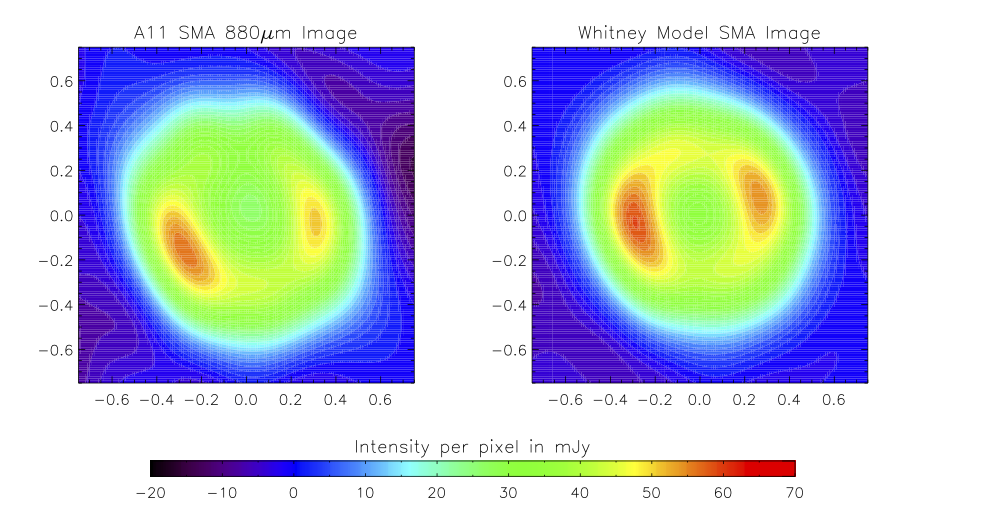}
\caption{Left: The 880$\mu$m SMA image of A11. Right: A large grain disk model with A11's best-fit input assumptions ($\delta$=10$^{-6}$, M$_{disk}$ = 0.006M$_{\sun}$, r$_{gap}$=36AU, etc.) produced by the Whitney code.}\label{fig8}
\end{figure}

\clearpage

\begin{figure}
\includegraphics[scale=0.8]{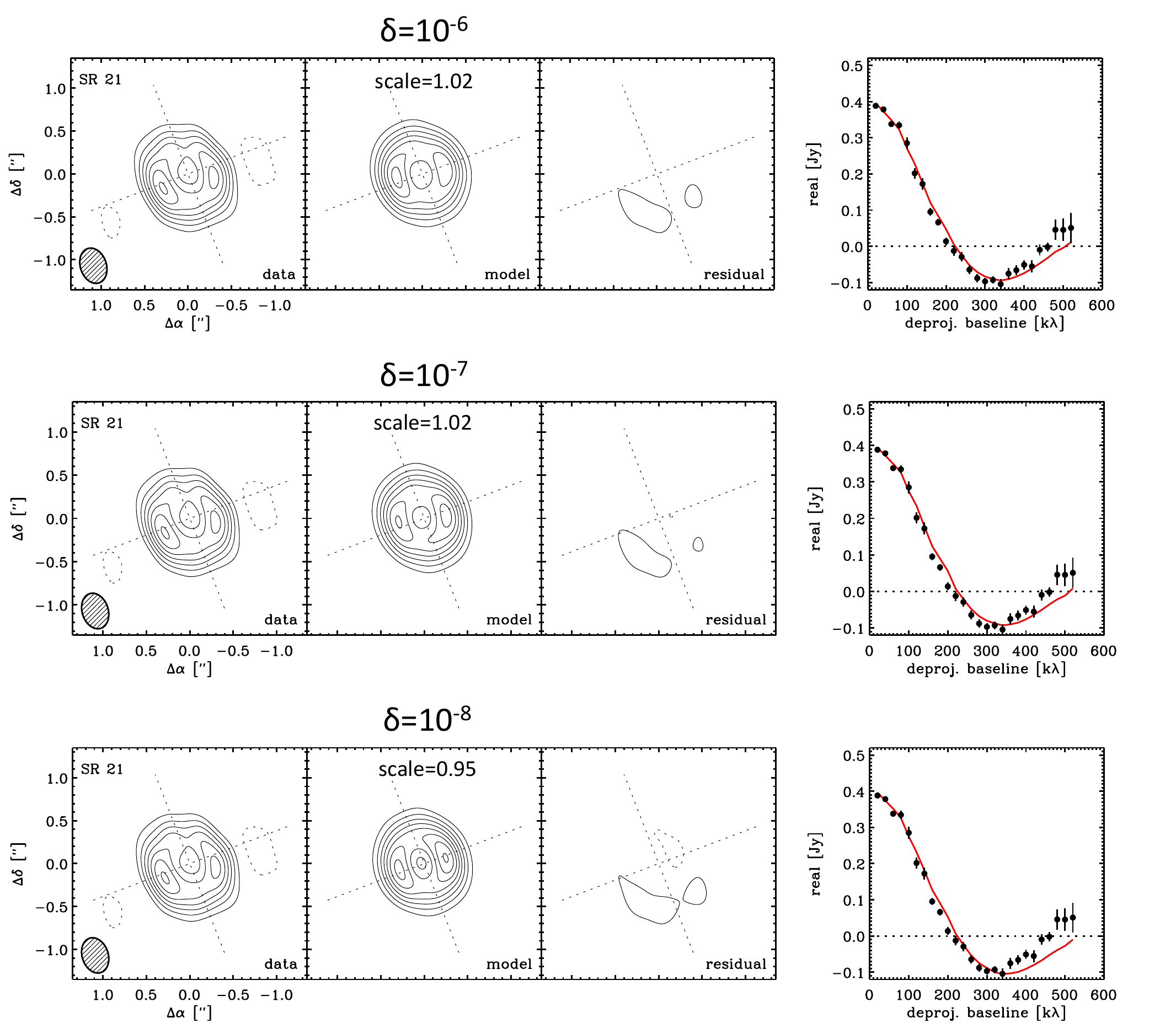}
\caption{Highly depleted ($\delta$=10$^{-8}$-10$^{-6}$) large grain disk models produced by the Whitney Code. From left to right, the panels are (a) the observed SR21 SMA image of A11 (b) the final Whitney model image, where the raw model output was sampled with the same spatial frequencies observed with the SMA (c) the residuals and (d) the model fit (in red) to the SMA visibilities. Input parameters were identical to those of A11, except that the depletion factor ($\delta$) for large grains inside of 36AU was varied. The depletion factor is given above each model row. A scale factor was also applied to each model image to equalize the model and SMA fluxes before subtraction from the observed image. This factor is given in panel (b) of each model. The minor discrepancies between modeled and observed fluxes can be easily explained by slight differences in input opacities. The bottom image is the same best fit model shown in Figure 7.} \label{fig9}
\end{figure}

\begin{figure}
\includegraphics[scale=0.8]{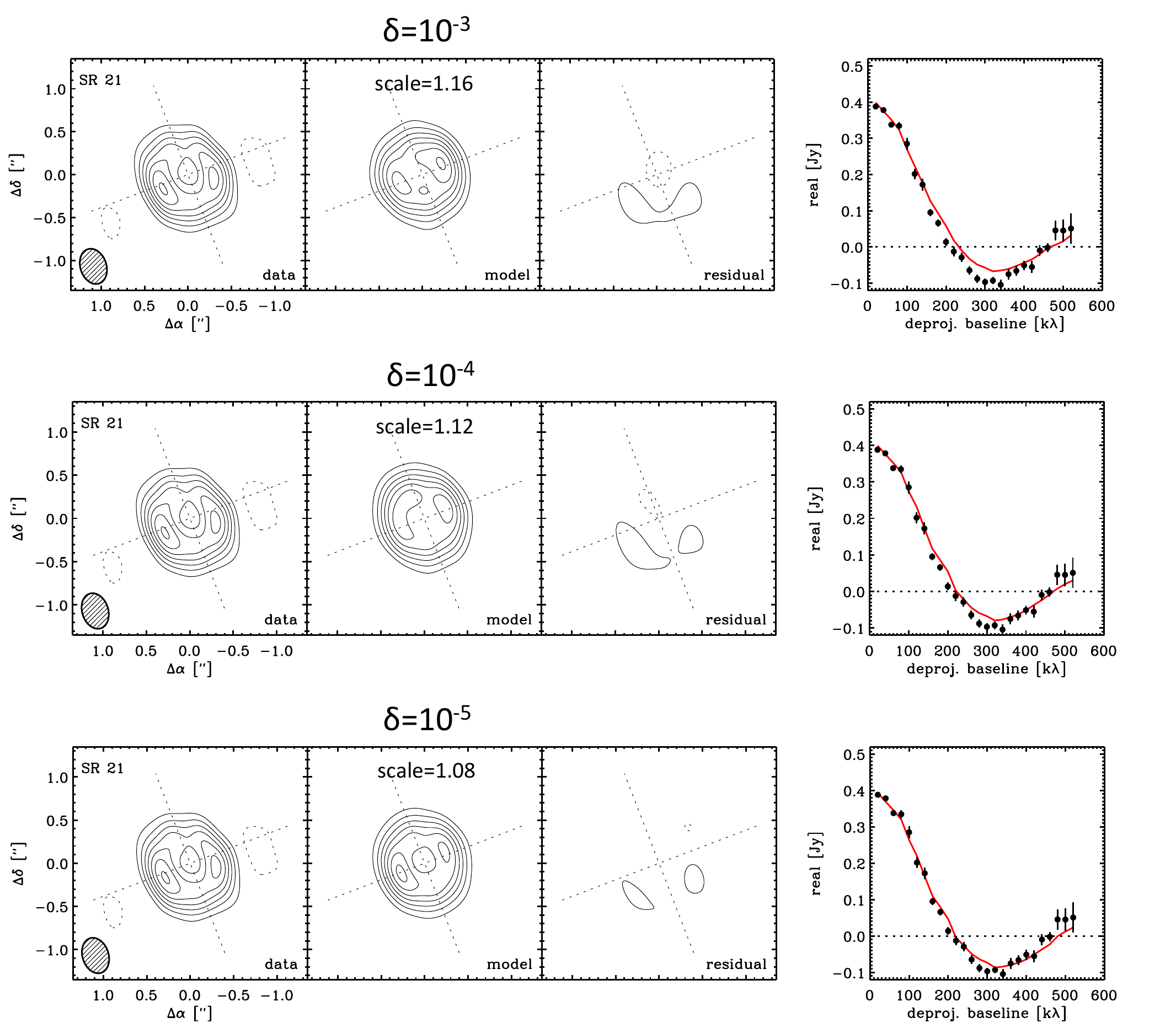}
\caption{Moderately depleted ($\delta$=10$^{-5}$-10$^{-3}$) large grain disk models produced by the Whitney Code. See Fig. 9 caption for details.} \label{fig10}
\end{figure} 

\begin{figure}
\includegraphics[scale=0.8]{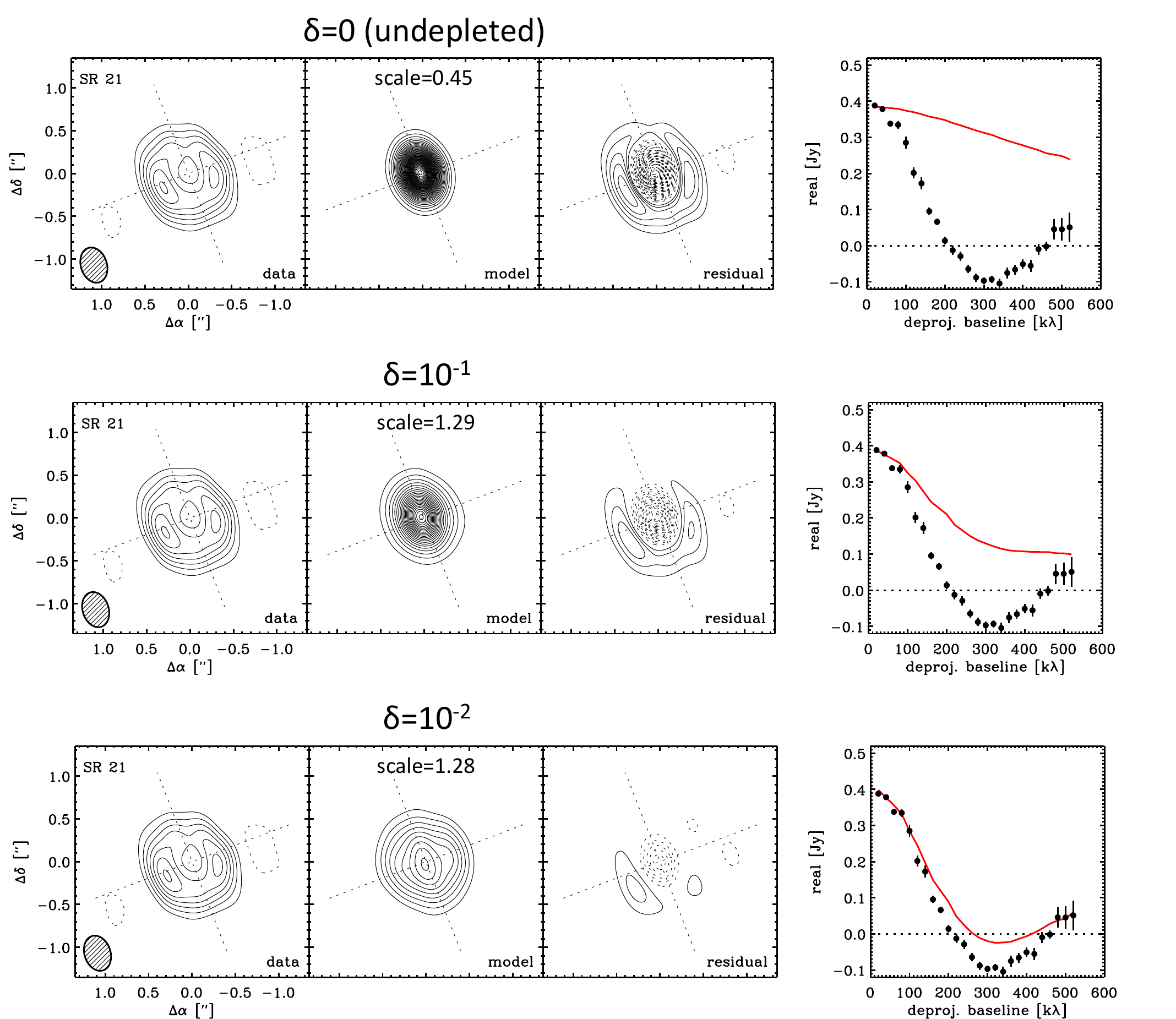}
\caption{Minimally depleted ($\delta$=10$^{-2}$-10$^{0}$) large grain disk models produced by the Whitney Code. See Fig. 9 caption for details.} \label{fig11}
\end{figure}

\begin{figure}
\includegraphics[scale=1]{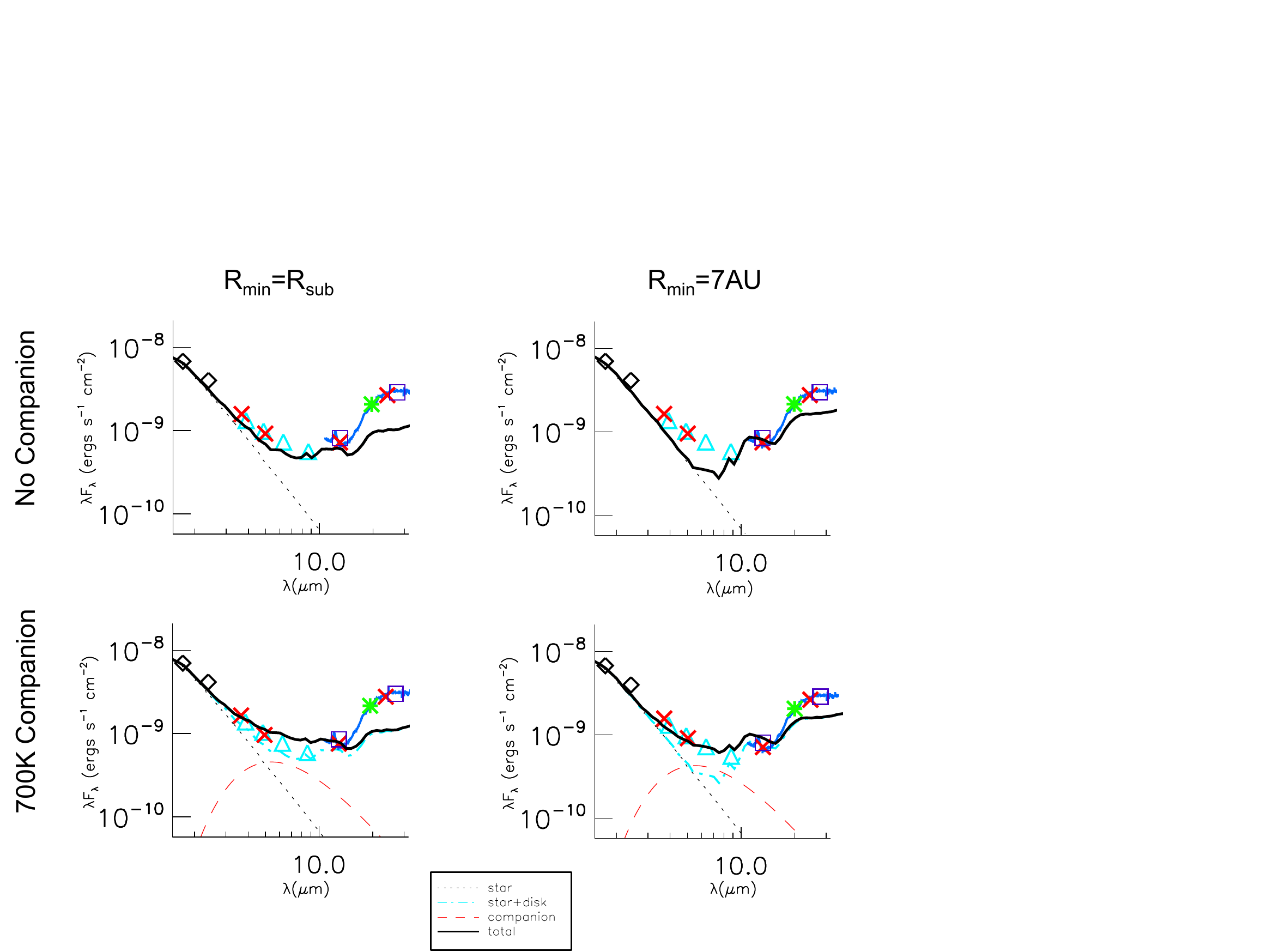}
\caption{Model NIR SEDs produced by otherwise identical models with and without (bottom and top panels respectively) the addition of a 700K blackbody companion to the SED and with either a 7AU inner disk truncation (right) or a disk extending inward to the sublimation radius (left).  These reveal that the $\lambda<$12$\mu$m NIR excess can be reproduced either by extending the disk inward to the sublimation radius (upper left) or by truncating the disk at a larger radius and adding in the NIR contribution of a 700K blackbody companion (lower right).}\label{fig12}
\end{figure}

\clearpage

\begin{figure}
\includegraphics[scale=1]{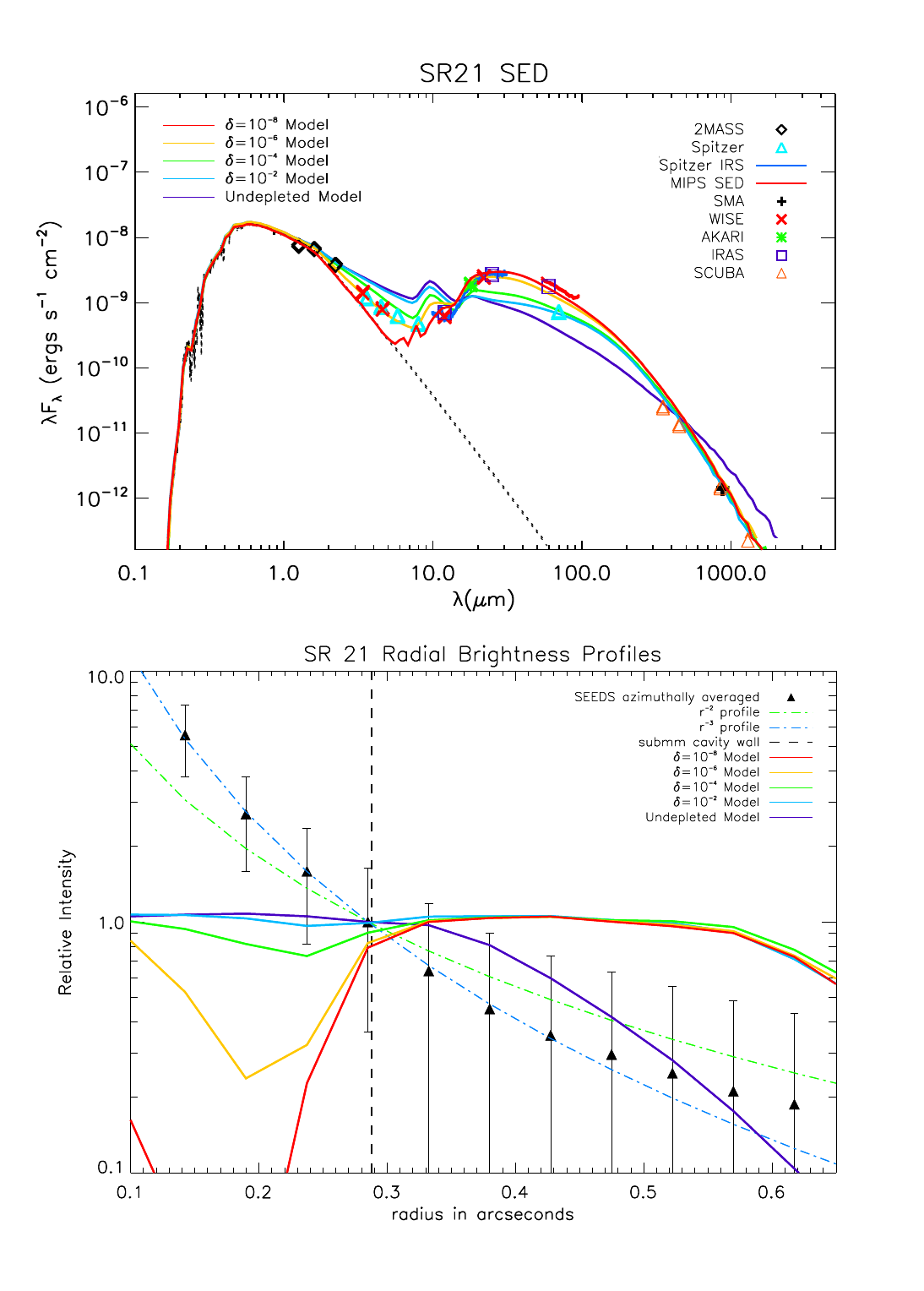}
\caption{Top: model SEDs of varying $r$$<$36AU depletion factors overplotted on the observed SR21 SED. Five ``Gapped" model runs of different depletion factors ($\delta$)are shown with solid colored lines, ranging from 10$^{-8}$ depletion in red to zero depletion in purple.  In the case of the heavily depleted $\delta$=10$^{-8}$ and $\delta$=10$^{-6}$ models, the NIR excess in the SED can be reproduced either by extending the disk inward to the sublimation radius or by adding the contribution of a 700K blackbody companion. This is discussed in detail in section 4.3.1 Bottom: H-band radial polarized intensity profiles for the same models overplotted on the observed profile. The flux of the undepleted model was fixed to the data at $r$=36AU, and all other models scaled by the same amount, in order to highlight changes in the radial profile at the cavity wall. None of the models provide a good fit to the data, revealing that the small grain disk is not only undepleted, its geometry is different from that of the large grain disk.}\label{fig13}
\end{figure}

\begin{figure}
\includegraphics{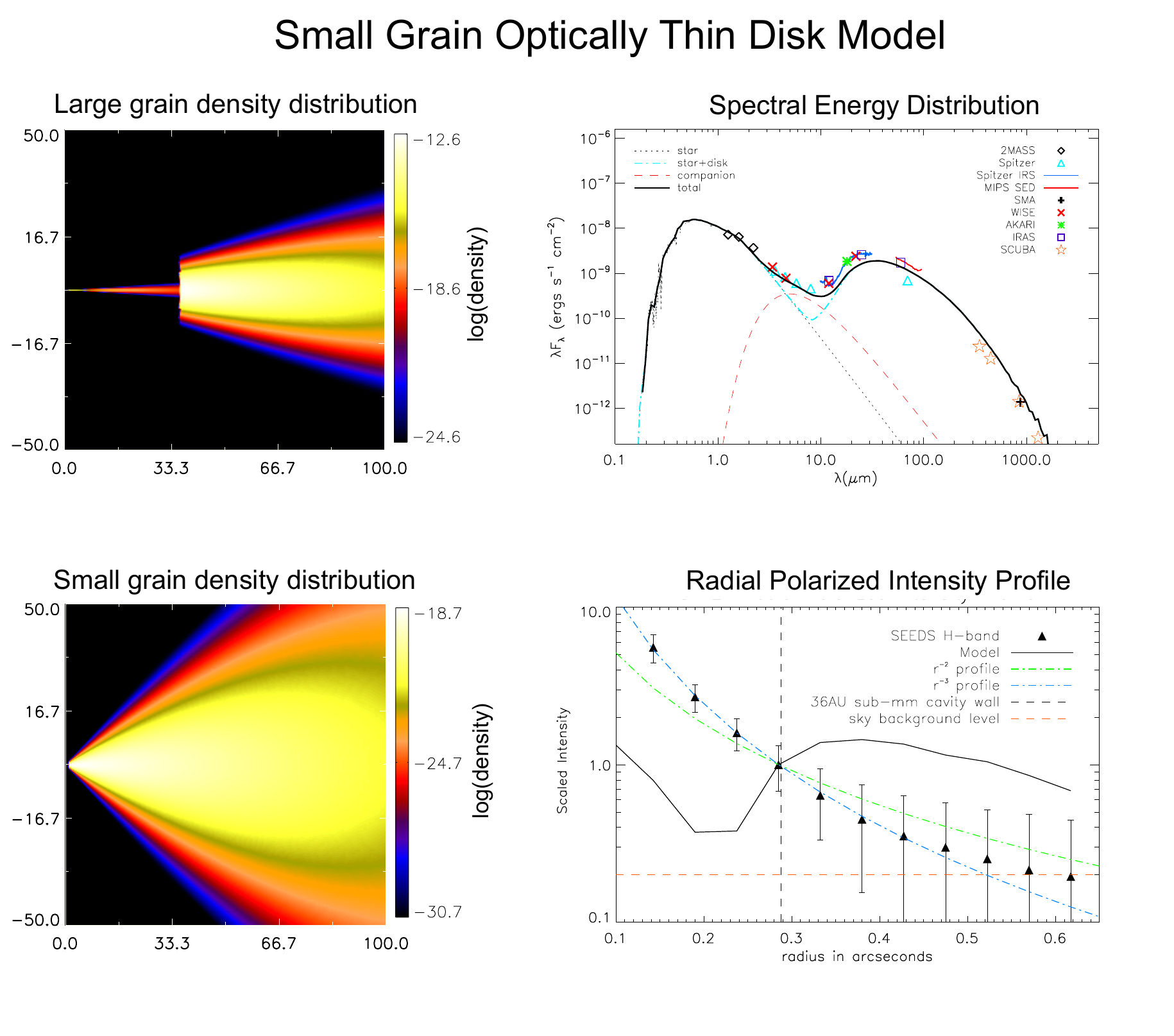}
\caption{Sample model outputs for a representative optically thin disk model. The lefthand panels represent the density of large grains (upper) and small grains (lower). The large grain model is identical to the A11 best-fit model described in Section 4.2 except that the dust wall at 36AU has been inflated by a factor of 2 in order to account for the lack of small grain contribution to the 12-20$/mu$m rise in the SED. The small grain model has an inner truncation at 7AU. The righthand panels provide model fits to the observed SED (upper) and H-band radial polarized intensity profile (lower). In this case, the lack of 10$\mu$m silicate emission is due to the very small amount of small grain material populating the inner disk. As revealed by the radial profile, an optically thick small grain disk alone cannot mask the discontinuity at the 36AU large grain dust wall. As justified in the text and shown in Figure 15, an underlying optically thick small grain layer is needed to mask this discontinuity, while this optically thin layer is required to reproduce the r$^{-3}$ radial polarized intensity profile.}\label{fig14}
\end{figure}

\begin{figure}
\includegraphics{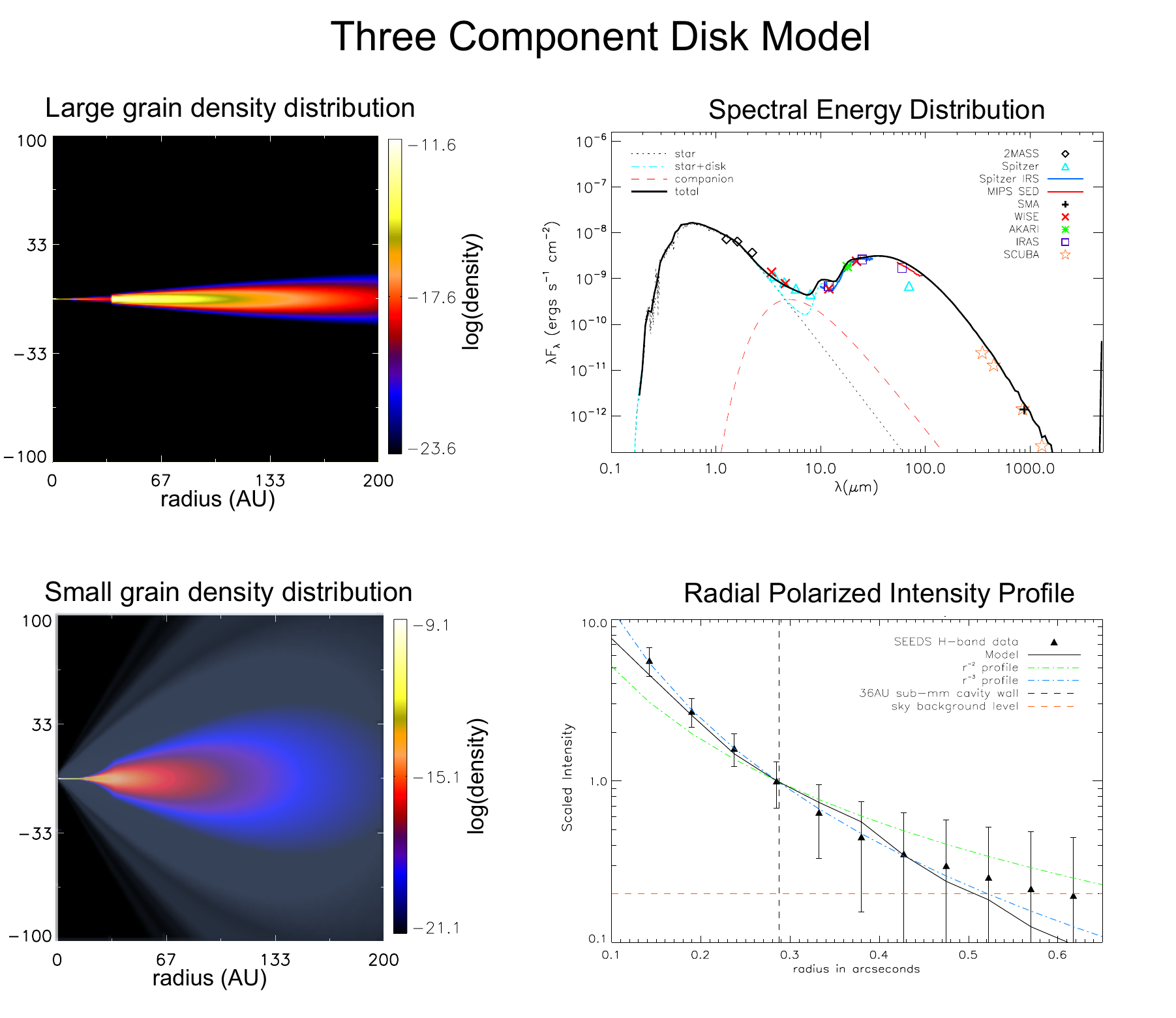}
\caption{Sample model outputs for a representative three component disk model. The lefthand panels represent the density of large grains (upper) and small grains (lower). The large grain model is identical to the A11 best-fit model described in Section 4.2.  The small grain disk consists of two components: (1) an underlying optically thick small grain disk of the `Steeply Curved' variety with a 7AU inner truncation and (2) an overlying vertically-extended  optically thin component that contributes thirty times more scattered light than the optically thick component.  The righthand panels provide model fits to the observed SED (upper) and H-band radial polarized intensity profile (lower). The presence of the optically thick small grain disk acts to mask the discontinuity at the large grain dust wall that was revealed in Figure 14. In this case, the 10$\mu$m silicate feature is small due to the small vertical extent of the optically thick small grain disk wall and the small amount of material populating the more vertically extended optically thin small grain component}\label{fig15}
\end{figure}

\begin{figure}
\includegraphics[scale=0.75]{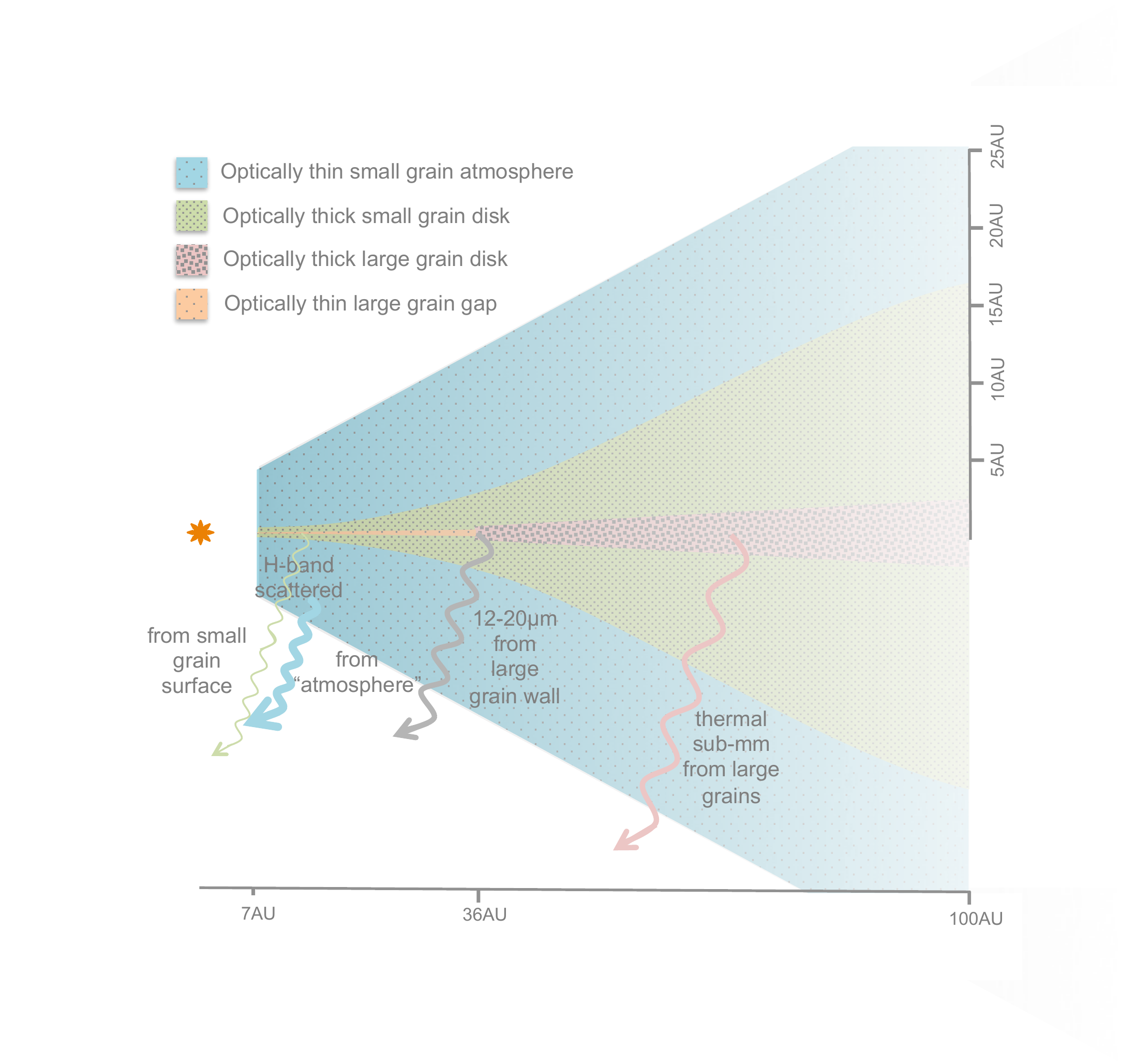}
\caption{A schematic view of the three component disk model that provides the best fit to our data. The vertical scale has been inflated by a factor of two relative to the horizontal. The origins of sub-mm light, H-band scattered light and the steep 12-20$\mu$m SED rise are indicated.}\label{fig16}
\end{figure}

\clearpage

\appendix
\section{Optically Thick Scattered Light Models}

In this appendix, we detail our attempts to derive a disk model in which the H-band scattered light emission originates from the surface of an optically thick small grain disk. Although we ultimately invoked an optically thin component in order to explain our data, this was not the most obvious solution, and more than 100 optically thick small grain scenarios were investigated before we arrived at the conclusion that the steepness of the observer H-band radial intensity profile is due to the effect of an optically thin small grain disk component. 

Due to the relative expense of full scattered light models, we have not explored the entirety of the parameter space for the optically thick small grain disk solution. We therefore describe our efforts in this appendix, in the hope that they will help to inform future H-band scattered light modeling efforts.

\subsection{Gapped and Differentially Depleted Models}
If the $r$$<$36AU gap is optically thin at all wavelengths, as has been assumed by most previous models, then sufficient flux is incident on the dust wall at 36AU to create a sharp rise in the SED from 12-20$\mu$m. We call the class of models with a depleted cavity that is optically thin in both large and small grains ``Gapped" models.  

Our initial working assumption, though perhaps naive, was that the observed SR21 cavity would also be seen in small grains. As discussed in section 4.3, this hypothesis provides an excellent fit to the multiwavelength SED. However, if the cavity size and depletion factor suggested by the sub-mm data is carried over to the small grains that scatter at H-band, the simulated H-band polarized intensity radial profiles produced by the Whitney code (Figure 17) suggest that we should have seen an observable discontinuity in the H-band scattered light PI profile at the $r$=36AU sub-mm wall. We see neither a slope nor a flux discontinuity at this radius, and can immediately exclude the simple scenario in which the large and small grains follow precisely the same distribution.  

The simplest scenario that allows us to maintain a similar distribution for large and small grains is one in which the two populations are both depleted inside of the cavity, but each to a different degree (a different $\delta$ value). Physically, this would seem to make sense, as the large grains concentrated at the midplane should be subject to different physical processes (perhaps most notably clearing by forming protoplanets) than the grains at the surface of the disk where H-band scattered light originates. Second generation small grains are also expected to be generated by the planet formation process \citep[e.g.,][]{Zuckerman:2001}, and so even if small grain depletion also occurs at the midplane, it could in some sense be compensated for. 

We began by exploring a variety of scenarios in which the small grains follow the same basic distribution as the large grains, are undepleted in the outer disk, but are depleted to a different degree than the large grains inside of the $r$$<$36AU cavity. Radial profiles for models following the same basic distribution as the large grain disk ($\alpha$=2.15, $\beta$=1.15), but with depletion values ranging from 10$^{-8}$ to zero (10$^{0}$) depletion, are shown overplotted on the observed H-band radial profile and on the observed SED in Figure 13. 

In moderate gap depletion cases ($\delta$=10$^{-3}$-10$^{-5}$), the flux drops sharply inside of the gap and falls all the way into the saturation radius, which is inconsistent with our observation of a continually inwardly increasing profile. In cases where the depletion is high enough to make the disk optically thin ($\delta$=10$^{-6}$-10$^{-8}$), the radial profile well inside the gap does increase sharply inward, however there is invariably a brightness discontinuity at the gap radius. This is due to the small amount of ``wall" that is tipped into the line of sight by the disk inclination, which creates an excess at this radius that is evident in the radial profile.  In low gap depletion cases ($\delta$=10$^{0}$-10$^{-2}$), the radial profiles are relatively smooth, but too flat.

The ``Gapped" models provide excellent fits to the SED and sub-mm imagery. However, the increased emitting area of the gap ``wall" invariably produces a local excess at 0$\farcs$3.  In order for such models to fit the observed radial profile, two things would need to occur. First, scattered light emission from the wall (whose existence is required by the SED fits) needs to be somehow suppressed to avoid a sharp discontinuity in scattered brightness. If the disk were perfectly face-on, it may be possible to hide the excess emitting area of the wall, but the range of inclination estimates from a variety of multiwavelength work (A11, this work, P08) suggest that this is unlikely. 

Second, they require that the optically thin grains inside of the gap and optically thick grains outside of the gap share a nearly identical radial PI profile despite very different physical conditions. Steep radial profiles can be produced by both optically thin and optically thick disks, but the apparent precise equality between the two makes this scenario seem somewhat contrived.  We therefore exclude the ``Gapped Disk" models as incompatible with the scattered light imagery except under the most contrived scenarios. 

\subsection{Scale Height Discontinuity Models}

We also consider scenarios that do not require the small grains to be depleted (and therefore optically thin) inside of the large grain gap. A step-like discontinuity in the small grain scale height could contribute to the steep SED rise, without requiring that the inner disk be depleted in small grains relative to the outer disk and allowing for identical (though discontinuous) distributions from inner to outer disk.  We call this class of models ``Scale Height Discontinuity" models, and the outputs for a representative model are shown in Figure 18.  A related class of models utilize a localized ``puff" in the large grain outer disk wall.  Inflated disk rims are often invoked to explain variable emission in disks \citep[e.g.,][]{Flaherty:2012}. Under such models, the high incident flux on a disk wall causes higher temperatures at that location and ``puffs up" the rim. If this were to occur at the 36AU large grain disk wall, it is conceivable that it would also inflate the small grain scale height at this radius. 

At the limit of a perfectly vertical step-like discontinuity and a face on disk, there is no break in the radial profile between inner and outer disk at H-band because the optical depth of these grains is large and the wall is obscured by the disk surface as viewed from face-on. However, as with the gapped disk models, it is difficult to mimic this situation in an inclined disk, even one with low inclination like SR21. These models thus solve the problem of requiring a different disk geometry inside and outside the gap however they also invariably produce a local excess at the gap radius that is not seen in the data, just as the ``Gapped" models did. 

In the case of puffed disk rims, the discontinuity that causes the puff is localized to the ``wall", so the disk soon settles back into its normal scale height distribution.  These ``puffs" may have a small radial extent, but again it is difficult to construct a situation in which they wouldn't be visible as an excess in the radial profile. On the other hand, the puffed disk rim shadows a portion of the disk immediately exterior to it from incoming stellar flux, where the area of the shadowed region depends on the height of the ``puff".  This local excess (``puff") followed by a local minimum (shadowed region) could be smoothed out in the polarized intensity radial profile if the total radial extent is much smaller than the diffraction limit at H-band, mimicking our observed smooth radial profile. This precise balance again seems somewhat contrived, so we consider the puffed rim explanation to be relatively unlikely, and we do not include any of these models in our final fits. 

Owing to their continuous density distribution, brightness discontinuities in the ``Scale Height Discontinuity" models are less drastic than those created by the ``Gapped" disk models, so they are more likely to remain unresolved in our observational data. Non-diffraction limited performance, the peculiar shape of the HiCIAO PSF and the relative insensitivity of polarized intensity emission to sharp disk features (due mainly to low overall flux) could all conceivably smooth out such discontinuities in the radial profile to a greater extent than convolution with a 0.06" Gaussian PSF would suggest. However, we were unable to construct a disk model in which the radial profile was smooth enough to be consistent with our data and error bars, so we exclude these models as relatively unlikely candidates. 

\subsection{Steeply Curved Models}
If we relax the assumption that the small grain disk needs to have precisely the same parameters inside and outside the gap, we find that there are cases where a turnover in the radial scale height exponent $\beta$ at the 36AU wall can reproduce the steep SED rise. This situation does not require small grain depletion, nor does it require a discontinuity at the large grain dust wall. In such cases, the SED rise coincides with the region just inside of 36AU where the disk is most steeply curved, and in that sense mimics a ``wall". From a physical standpoint, this transition from a large to a smaller radial scale height exponent at the large grain dust wall would need to be explained somehow by the dearth of large grains at the disk midplane in this region or the physical mechanism causing disk clearing, which we do not endeavor to do here. We call this class of models ``Steeply Curved" models, and outputs for a representative model are shown in Figure 19.  

Under the third well-fitting SED scenario, a steep disk curvature (high $\beta$) in the inner disk mimics a ``wall" at the transition to a shallower outer disk profile. Without a physical discontinuity in the small grain surface at 36AU, these models are also free of discontinuity in their radial polarized intensity profiles.  However, all ``Steeply Curved" models also produce flat ($r$$^{0}$) radial profiles, which our observations argue against.  This is because the geometric emitting area per pixel increases as the disk becomes increasingly flared, so pixels represent a larger amount of disk surface as you move outward. This mitigates the $r$$^{-2}$ effect of incident stellar radiation, and makes the flux in the disk extremely flat. 

We find that the surface brightness of the disk is relatively insensitive to the radial density exponent $\alpha$ in our high-$\beta$ models, even for very strongly inwardly peaked density profiles (e.g., $r$$^{-10}$). These models still show a flattening of the brightness profile in steeply curved regions that our data rule out. We therefore exclude these high-$\beta$ undepleted scenarios as poor fits to the H-band data.  

We find that the steepness of the radial profile significantly limits the disk geometry, particularly in the optically thick case. We tried more than 30 iterations of smooth optically thick disks with a wide range of $\alpha$, $\beta$ and $\alpha$-$\beta$ (the radial dependence of the surface density) values, and were unable to create a steep $r$$^{-3}$ profile.  

The only optically thick disks we found that reproduce an $r$$^{-3}$ radial polarized intensity profile are geometrically flat. As the geometrically flat case requires that the scale height of the disk be smaller than the radius of the star, it is impossible to create one in which the large grain disk required to reproduce the sub-mm flux does not stick up above the small grain disk. It is difficult to imagine a situation that would reproduce this, so we exclude these models on physical grounds. 

 \subsection{Prospects for Envelope Component}

Since SR21 may still be weakly accreting (the upper limit of 10$^{-8.84}$ on the accretion rate is relatively high compared to other disks), we also explored the possibility that an envelope component is contributing to the flux in the interior of the gap.  To provide observational constraints on the amount of circumstellar material, SR21 was observed on 2011 March 29 with SIRPOL \citep{Kandori:2006}, a JHKs-simultaneous imaging polarimeter that is mounted on the Infrared Survey Facility (IRSF) 1.4m telescope in South Africa.  We conducted aperture polarimetry on this data and the results are summarized in Table 3.  Our data reveal total polarization levels that are consistent with the range of total polarization observed for YSOs \citep{Pereyra:2009}.  We have not corrected the SR21 aperture polarization values reported in Table 3 for interstellar polarization, which could be significant due to the high foreground extinction (A$_{V}$=6.3).  Moreover, because the observations were conducted in defocus mode to avoid saturation, the A and B components of SR21 (separated by 6") overlap in our aperture polarimetry, preventing us from using SR21B to determine how much of the measured polarization is due to foreground dust and how much is local to SR21A and/or B.  There are no other nearby objects in the field suitable for use as field star interstellar polarization probes.  

If we assume that the level of interstellar polarization and contamination from SR21B is minimal, we can use the measured polarization Position Angles (PAs) at each band to probe of the optical depth of the disk, as optically thick disks show polarization PAs oriented perpendicular to the PA of the disk major axis, while the polarization PA and major axis PAs of optically thin disks are oriented parallel to one another \citep{Whitney:1992,Pereyra:2009}. The disk major axis PA (86$\pm$11$^{\circ}$, this work; 100$^{\circ}$, A11) is roughly perpendicular to the measured polarization orientations at all three bands, suggesting that the disk may be optically thick in the NIR. 

However, there are several caveats to be made. First, the 42$^{\circ}$ counterclockwise migration in the disk PA derived from isophotal fits (and described in section 3.1) suggest that an average PA is not a particularly good measure of the overall disk orientation. Secondly, these results depend on the robustness of our assumption that the interstellar polarization and contamination from SR21B is small.  Finally, and most importantly, we have managed to eliminate all of the optically thick small grain dust disk scenarios that we have investigated thus far.  In fact, we find that optically thin disks provide a much better fit to the H-band radial profile, as described in Section 4.4, although we cannot rule out, and in fact require, an additional contribution from an underlying optically thick small grain disk component.

\begin{table} [ht]
\caption{SIRPOL Aperture Polarimetry}
\centering
\begin{tabular} {c c c}
\hline\hline
Band & Polarization & PA\\
 & (\%) & $^{\circ}$\\
\hline
J & 2.32$\pm$ 0.17 & 21.5$\pm$2.1 \\
H & 1.01$\pm$0.05  & 18.0$\pm$1.3\\
Ks & 0.52$\pm$ 0.10 & 179.2$\pm$5.3\\
\hline
\end{tabular}
\label{table:inputs}
\end{table}

\begin{figure}
\includegraphics[scale=1]{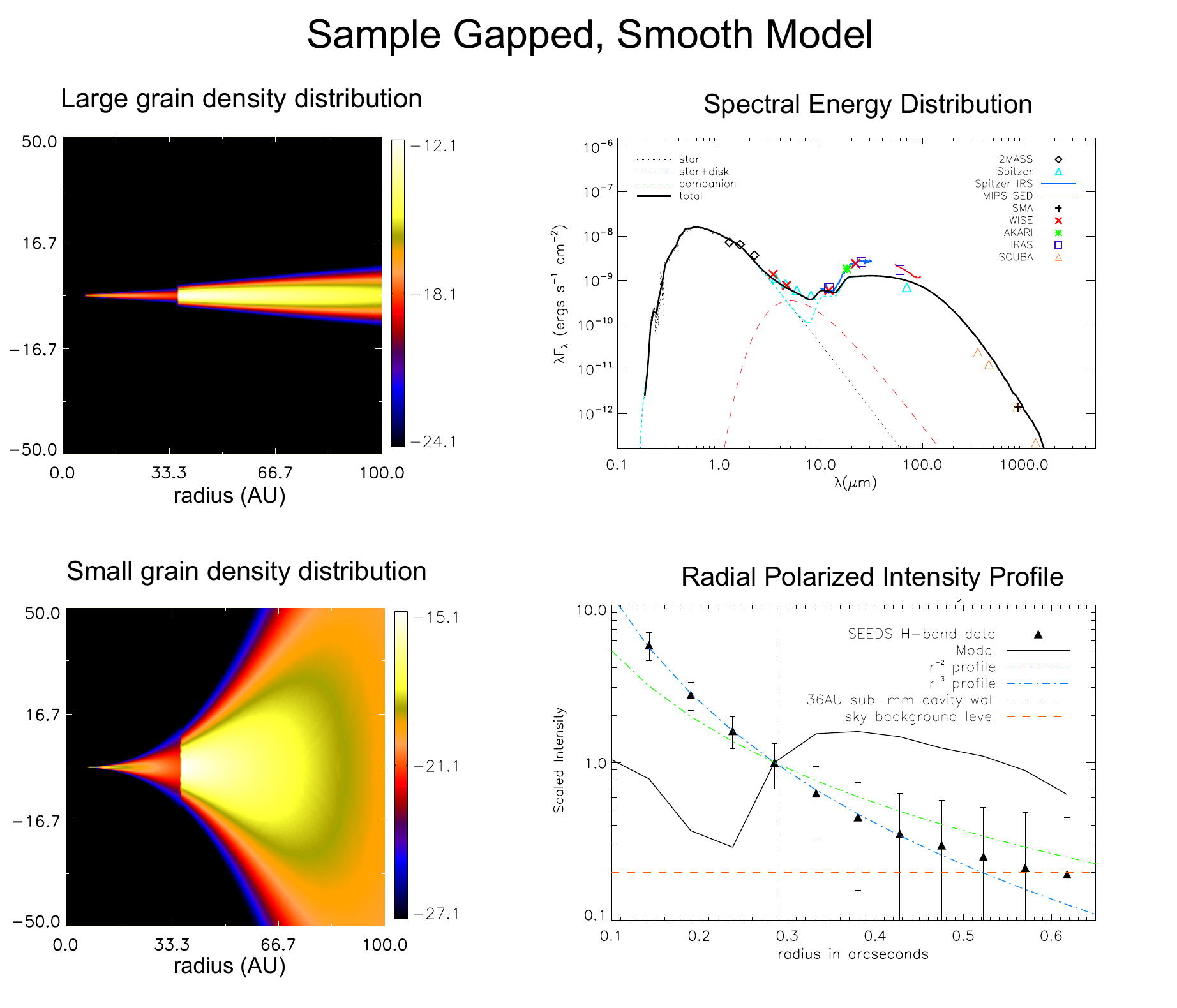}
\caption{Sample model outputs for a representative gapped disk model. The lefthand panels represent the density of large grains (upper) and small grains (lower). The large grain model is identical to the A11 best-fit model described in Section 4.2. The small grain disk, like the large grain disk, is depleted by a factor of 10$^{-6}$ interior to 36AU, and it truncated at 7AU. The righthand panels provide model fits to the observed SED (upper) and H-band radial polarized intensity profile (lower). Although the small grain surface layer is continuous under this model, the heavy small grain depletion interior to 36AU allows a significant amount of flux to be indecent on the wall, creating a discontinuity in the H-band polarized intensity profile.}\label{fig17}
\end{figure}

\begin{figure}
\includegraphics[scale=1]{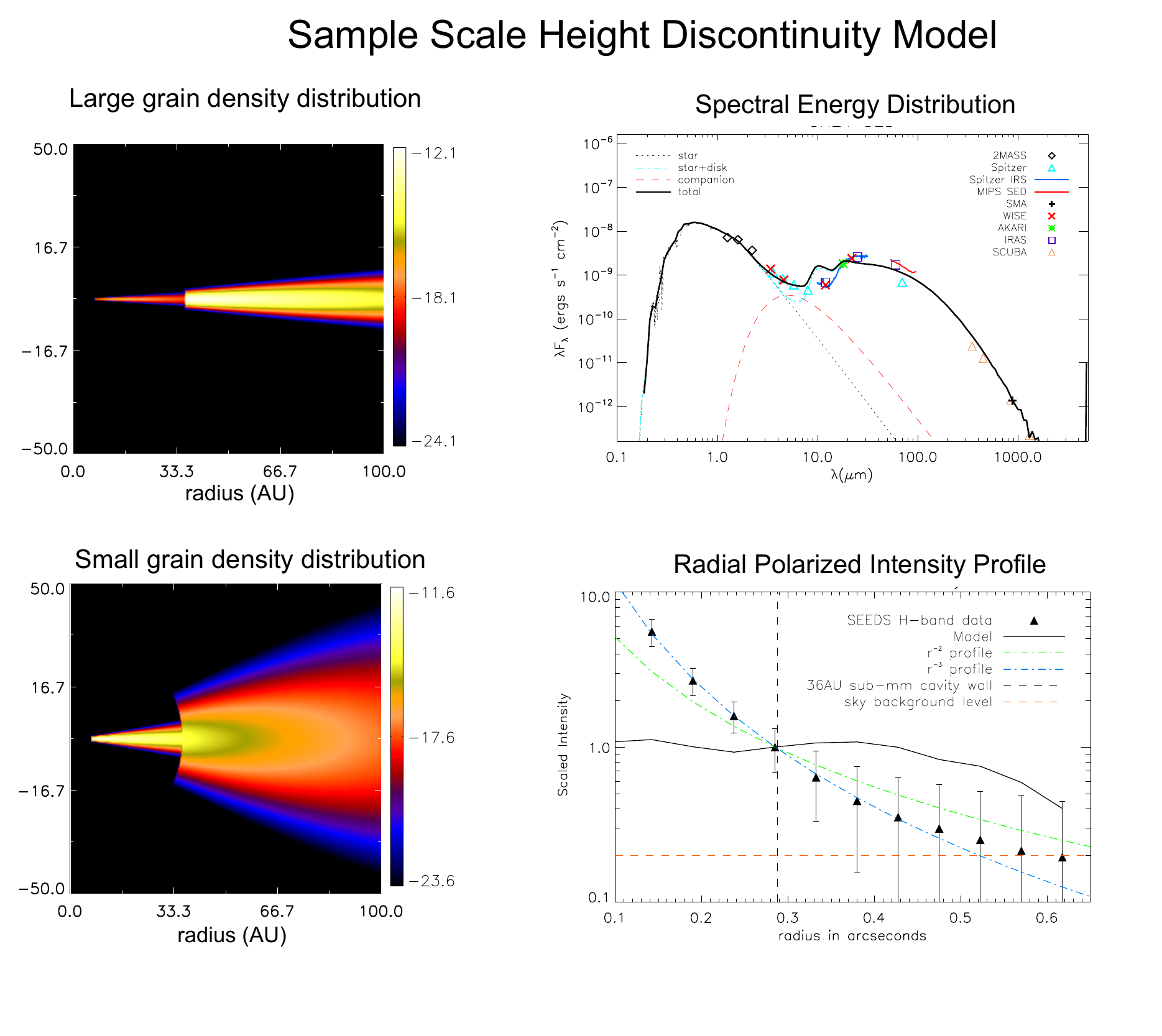}
\caption{Sample model outputs for a representative undepleted small grain disk with a scale height discontinuity at the 36AU large grain dust wall. The lefthand panels represent the density of large grains (upper) and small grains (lower). The large grain model is identical to the A11 best-fit model described in Section 4.2. The small grain disk undepleted, with a scale height discontinuity at 36AU and an inner truncation at 7AU. The righthand panels provide model fits to the observed SED (upper) and H-band radial polarized intensity profile (lower). The large amount of small grain material in the inner disk overproduces the 10$\mu$m flux, but provides an otherwise good fit to the SED. In this case the discontinuity in the radial profile at 36AU is smaller than in the gapped models due to the relatively smaller amount of direct stellar flux incident on the small grain dust wall. However, the steep r$^{-3}$ profile is not reproduced in this case, nor was it with any of the other radial density ($\alpha$) and scale height ($\beta$) exponents investigated for such scale height discontinuity models.}\label{fig18}
\end{figure}

\begin{figure}
\includegraphics[scale=1]{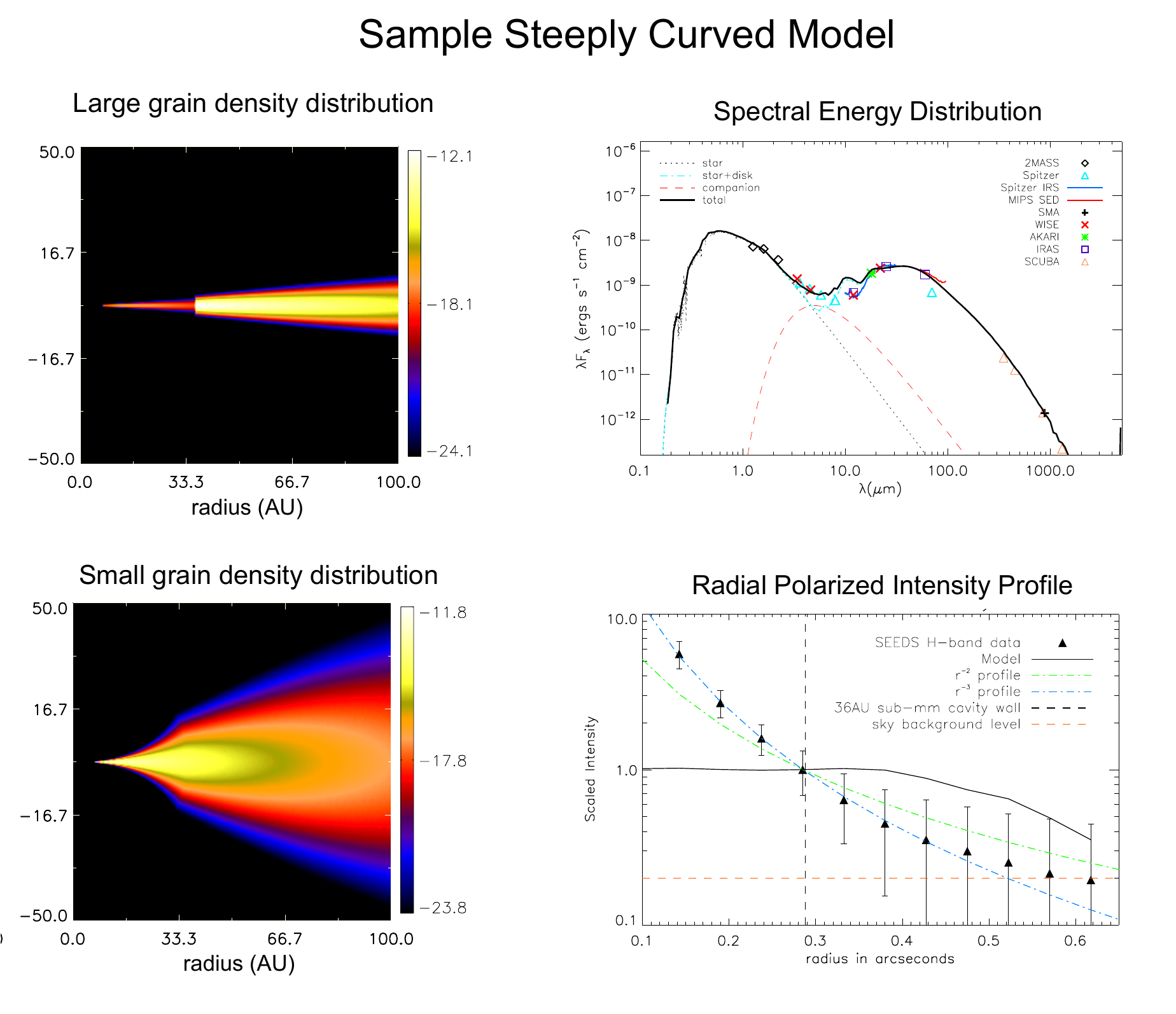}
\caption{Sample model outputs for a representative model in which the region of the sub-mm cavity is more steeply curved in the small grain disk. The lefthand panels represent the density of large grains (upper) and small grains (lower). The large grain model is identical to the A11 best-fit model described in Section 4.2. The small grain model is undepleted interior to 36AU and truncated at 7AU. The scale height exponent $\beta$ was increased from 1.15 in the outer disk to 2.5 interior to 36AU, allowing for a steeper curvature. The righthand panels provide model fits to the observed SED (upper) and H-band radial polarized intensity profile (lower). This small grain distribution completely masks any discontinuity at the 36AU dust wall created by an H-band contribution from the large grain dust prescription, however the steep inner disk carries a flat radial profile, which is inconsistent with the observed r$^{-3}$ profile.}\label{fig19}
\end{figure}

\bibliographystyle{apj}
\bibliography{SR21_revise}

\end{document}